\documentclass[aps,prc,preprint,amsmath,amssymb,showpacs,preprintnumbers,superscriptaddress]{revtex4-2}
\usepackage{CJK}
\usepackage{graphicx}
\usepackage{dcolumn}
\usepackage{bm}
\usepackage{bbm}
\usepackage{color}
\usepackage{hyperref}
\usepackage{booktabs}
\usepackage{amssymb}

\allowdisplaybreaks

\begin{document}

\title{Configuration-interaction time-dependent density functional theory for nuclear dynamics}

\author{Y. P. Wang}
\affiliation{State Key Laboratory of Nuclear Physics and Technology, School of Physics, Peking University, Beijing 100871, China}
\author{B. Li}
\affiliation{State Key Laboratory of Nuclear Physics and Technology, School of Physics, Peking University, Beijing 100871, China}
\author{D. Vretenar}
\email{vretenar@phy.hr}
\affiliation{Physics Department, Faculty of Science, University of Zagreb, 10000 Zagreb, Croatia}
\affiliation{State Key Laboratory of Nuclear Physics and Technology, School of Physics, Peking University, Beijing 100871, China}
\author{T. Nik\v{s}i\'c}
\affiliation{Physics Department, Faculty of Science, University of Zagreb, 10000 Zagreb, Croatia}
\affiliation{State Key Laboratory of Nuclear Physics and Technology, School of Physics, Peking University, Beijing 100871, China}
\author{P. W. Zhao}
\email{pwzhao@pku.edu.cn}
\affiliation{State Key Laboratory of Nuclear Physics and Technology, School of Physics, Peking University, Beijing 100871, China}
\author{J. Meng}
\email{mengj@pku.edu.cn}
\affiliation{State Key Laboratory of Nuclear Physics and Technology, School of Physics, Peking University, Beijing 100871, China}

\date{\today}

\begin{abstract}

A configuration-interaction time-dependent density functional theory (CI-TDDFT) for nuclear dynamics is developed. In this framework, the correlated nuclear many-body wave function is expanded in terms of time-dependent many-particle configurations built from a common set of orthonormal single-particle states. The equations of motion for both the expansion coefficients and the single-particle states are derived self-consistently using the Dirac-Frenkel time-dependent variational principle. This formulation extends conventional time-dependent density functional theory (TDDFT) by incorporating configuration mixing and beyond-mean-field correlations, while preserving energy and particle-number conservation.

As an illustrative application, the method is implemented using the relativistic point-coupling functional PC-PK1 in the particle-hole channel and a monopole pairing interaction in the particle-particle channel, and is applied to the study of isoscalar giant monopole resonance in $^{58}$Ni and $^{60}$Ni.
Numerical tests show that both the total energy and particle number are conserved, with relative deviations within $4\times 10^{-4}$ during the time evolution.
Compared with conventional TDDFT, CI-TDDFT yields broader strength distributions for giant monopole resonances while keeping the main peak positions close to those from TDDFT. This broadening is associated with configuration mixing in the valence space and suggests a coupling of the monopole oscillation to additional collective degrees of freedom. These results demonstrate the potential of CI-TDDFT as a quantum, microscopic beyond-mean-field framework for nuclear dynamics.
\end{abstract}

\maketitle


\newpage

\section{Introduction}

Nuclear time-dependent density functional theory (TDDFT) is a central microscopic framework for describing nuclear dynamics. Starting from a nuclear density functional, TDDFT models the dynamics of a complex many-body system using a product-type wave function that describes the time evolution of independent nucleons in a self-consistent mean field \cite{Nakatsukasa2016RMP,Simenel2018PPNP,Negele1982RMP,Ren2020PLB}. This approach enables the microscopic and computationally feasible study of complex nuclear dynamical phenomena across wide regions of the nuclear chart. Over the past few decades, TDDFT has been successfully applied to a diverse range of dynamical processes, including small-amplitude collective vibrations \cite{Blocki1979PLB,Vretenar1995NPA,Ring1996NPA,Reinhard2007EPJA,GengJ2025PRC}, as well as large-amplitude phenomena such as fission \cite{Bulgac2016PRL,Ren2022PRL,LiB2023PRC2}, fusion \cite{Umar2006PRC,Simenel2013PRC,RenZX2020PRC}, and multinucleon transfer \cite{Simenel2010PRL,Wu2019PRC,Zhang2024PRC,LiB2024PRC,ZhangDD2025PLB,YangYC2026PRC}.

Despite its broad applicability, TDDFT has certain limitations stemming from its mean-field character. It is well known that TDDFT reduces to the random-phase approximation (RPA) in the small-amplitude limit, describing collective vibrations primarily in terms of coherent one-particle-one-hole (1p-1h) configurations, while higher-order (2p-2h, 3p-3h, ..., np-nh) configurations are not explicitly included \cite{Ring2004The}. Consequently, TDDFT systematically underestimates the spreading widths of giant resonances \cite{Scamps2013PRC,Harakeh2001GiantResonances}. For large-amplitude processes such as fission, TDDFT describes the classical evolution of independent nucleons through the propagation of the one-body density. Although this framework naturally incorporates one-body dissipation, it lacks quantum fluctuations in the collective space and therefore cannot reproduce the widths of charge or mass distributions \cite{Bulgac2016PRL,Ren2022PRL}. Moreover, TDDFT is not directly applicable to quantum many-body tunneling processes, such as fission dynamics before the saddle point \cite{Bulgac2016PRL,Ren2022PRL} and fusion reactions at energies below the Coulomb barrier \cite{Umar2006PRC,Simenel2013PRC,RenZX2020PRC}.

These limitations have motivated several beyond-mean-field extensions of the conventional TDDFT framework, including the time-dependent density-matrix (TDDM) theory \cite{Wang1985AnnPhys,Assie2009PRL,Wen2018PRC,Tohyama2020FrontPhys}, the time-dependent random-phase approximation (TDRPA) method \cite{Balian1981PRL,Balian1984PLB,Simenel2011PRL,Williams2018PRL,Simenel2020PRL}, stochastic time-dependent Hartree-Fock (TDHF) \cite{Ayik2008PLB,Lacroix2014EPJA,Tanimura2017PRL}, and the generalized time-dependent generator-coordinate method (gd-TDGCM) \cite{LiB2023PRC,Marevic2023PRC,Marevic2024EPJA,Li2024FOP,Li2025PRC}. 
Based on the Bogoliubov-Born-Green-Kirkwood-Yvon (BBGKY) hierarchy \cite{Born1946PRS,Bogoliubov1946JPhysUSSR,Kirkwood1946JCP}, TDDM couples the time evolution of the one-body density matrix to that of the two-body correlation matrix, thereby incorporating two-body correlations beyond the mean-field level \cite{Tohyama2020FrontPhys}.
However, its practical applications are limited by the large dimensionality of the two-body correlation matrix and typically require additional truncations, for example by retaining only correlations between time-reversed pairs \cite{Assie2009PRL}.
TDRPA, derived from the Balian–Vénéroni variational principle \cite{Balian1981PRL,Balian1984PLB}, provides a practical way to evaluate fluctuations of one-body observables around a TDDFT trajectory, but it remains essentially a small-amplitude treatment around the mean-field evolution \cite{Simenel2011PRL,Williams2018PRL,Simenel2020PRL}. Stochastic TDHF incorporates one-body fluctuations by replacing a single deterministic trajectory with an ensemble of mean-field trajectories generated from fluctuating initial one-body density matrices. However, its predictive power depends on the sampling prescription, and the resulting trajectories are averaged statistically rather than coherently \cite{Ayik2008PLB,Lacroix2014EPJA,Tanimura2017PRL}.

Among these beyond-mean-field extensions, the gd-TDGCM is particularly appealing because it combines advantages of both TDDFT and TDGCM by incorporating dissipation effects and quantum fluctuations in collective space. The gd-TDGCM was originally introduced in Ref.~\cite{Reinhard1983NPA}, and has recently been developed within both nonrelativistic \cite{Marevic2023PRC,Marevic2024EPJA} and relativistic \cite{LiB2023PRC,Li2024FOP,Li2025PRC} density functional theory (DFT) frameworks. In this approach, the correlated nuclear wave function is expanded in terms of time-dependent many-body generator states. Typically, these generator states are initialized from deformation-constrained self-consistent mean-field calculations and form a set of nonorthogonal, generally overcomplete basis states. For practical reasons, these states are usually assumed to propagate independently along TDDFT trajectories, while only the expansion coefficients are determined by the variational principle. Consequently, the feedback of configuration mixing on the evolution of the basis states is neglected, and strict energy conservation is no longer guaranteed. Although this issue could in principle be resolved by a fully variational evolution of both the basis states and the expansion coefficients, such a framework is considerably more complicated for nonorthogonal generator states.

This motivates us to develop an alternative framework in which the many-body basis is orthogonal and both the basis states and the expansion coefficients are treated on the same variational footing. Many-particle configurations (MPCs)—constructed from a common set of single-particle states by generating particle-hole excitations (1p-1h, 2p-2h, …, np-nh) with respect to a reference configuration—provide such an orthogonal basis. In static nuclear structure calculations, configuration-interaction techniques based on particle-hole or quasiparticle excitations have been successfully employed, for instance in relativistic configuration-interaction density functional theory (ReCD) \cite{Zhao2016PRC_CIPDFT,Wang2022PRC_CIPDFT,WangYK2024PLB,WangYK2024PLB2,WangYK2024SB,QuT2025PRC}, the configuration-interaction relativistic Hartree-Fock (CI-RHF) model \cite{Liu2025CPC,PengY2025CPC}, and the shell-model-like approach (SLAP) based on the relativistic density functional theory (RDFT-SLAP) \cite{MengJ2006FPC,ShiZ2018PRC,WangYP2023PLB,WangYP2024PRL,XuFF2024PRL}. For dynamical processes, related multiconfigurational time-dependent methods have become standard and highly successful tools in quantum chemistry \cite{Meyer1990MCTDH,Beck2000PhysRep,WangThoss2003MLMCTDH,Manthe2008MLMCTDH,Sato2013PRA}. These methods and their extensions have been applied successfully to a range of molecular dynamical phenomena, including photodissociation  \cite{Manthe1992JCP_NOCl,Giri2011JCP_NH3,Westermann2011JCP_CH3IHost,Han2022JCTC_MCTDH}, photoabsorption \cite{Worth1996JCP_Pyrazine,Beck2000PhysRep, Vendrell2011JCP_MLMCTDH,Meng2014JCP_CH2OO}, and molecule-surface scattering \cite{Ehara1996JCP_MCTDHSurface,Heitz2001JCP_N2LiF,Crespos2006JCP_H2Pt,Meng2017JCP_COCuSurface,Meng2021JCTC_COCuSurface,Song2022JPCA_H2OCu}. In nuclear systems, however, the complexity of the nuclear force and the DFT framework makes such developments considerably more challenging, and time-dependent configuration-interaction approaches based on orthogonal MPCs remain largely unexplored.

In this work, we develop a configuration-interaction time-dependent density functional theory (CI-TDDFT) for nuclear dynamics and present its first application to the isoscalar giant monopole resonance (ISGMR) in $^{58}$Ni and $^{60}$Ni. The ISGMR serves as an ideal initial test case because standard TDDFT can reproduce its centroid energy reasonably well, whereas the spreading width requires correlations beyond the 1p-1h level \cite{Scamps2013PRC,Harakeh2001GiantResonances}. The proposed CI-TDDFT method offers several potential advantages. First, it provides a self-consistent framework for incorporating one-body, two-body, and more generally many-body correlations. Second, it is systematically improvable in principle, as convergence can be approached by increasing the number of MPCs included in the expansion. Third, the total energy is conserved as a consequence of the fully variational evolution of both the expansion coefficients and the single-particle states. Finally, the chosen form of the many-body wave function ensures particle-number conservation even in the presence of pairing correlations. 
 
The article is organized as follows. 
In Sec.~\ref{Theory}, the theoretical framework is developed. 
In Sec.~\ref{Numerical-details}, the numerical details for solving the equations of motion for the expansion coefficients and single-particle states are presented.
In Sec.~\ref{Numerical-test}, numerical tests of energy and particle-number conservation are performed. 
An illustrative application to the ISGMR in $^{58}$Ni and $^{60}$Ni is discussed in Sec.~\ref{Results-and-discussions}. 
Finally, the main results are summarized and a brief outlook for future studies is presented in Sec.~\ref{summary-and-outlook}.

\section{Theoretical framework}\label{Theory}

\subsection{Nuclear many-body wave function}

In CI-TDDFT, the correlated nuclear wave function is expanded in terms of time-dependent many-particle configurations $|\text{MPC}_I(t)\rangle$:
\begin{equation}
\label{wave-function}
|\Psi(t)\rangle = \sum_I C_I(t) |\text{MPC}_I(t)\rangle,
\end{equation}
where $\{C_I(t)\}$ is the set of the expansion coefficients.

Each configuration $|\text{MPC}_I(t)\rangle$ is constructed from a common set of $N_{\text{spl}}$ orthonormal single-particle states as follows
\begin{equation}
|\text{MPC}_I(t)\rangle = \prod_{i=1}^{N_{\text{spl}}^{\text{n}}} \big(c_{i,\text{n}}^{\dagger}\big)^{I_{i,{\text{n}}}}\prod_{j=1}^{N_{\text{spl}}^{\text{p}}}
\big(c_{j,\text{p}}^{\dagger}\big)^{I_{j,{\text{p}}}}|-\rangle,
\end{equation}
with $|-\rangle$ denoting the vacuum state, and $N_{\text{spl}}=N_{\text{spl}}^{\text{n}}+N_{\text{spl}}^{\text{p}}$. 
Here, $c_{i,\tau}^{\dagger}$ creates a nucleon in the single-particle state $|\phi_{i,\tau}\rangle$, and $\tau = \text{n}$ (neutron) or $\text{p}$ (proton). 
The occupation of each state is specified by the integer array $\{I_{i,{\tau}} = 0,1\}$, so that $\sum_{i=1}^{N_{\text{spl}}^{\text{n}}} I_{i,{\text{n}}} = N$ and $\sum_{i=1}^{N_{\text{spl}}^{\text{p}}} I_{i,{\text{p}}} = Z$ for a system with $N$ neutrons and $Z$ protons.

\begin{figure*}[htbp!]
   \centering
   \includegraphics[width=0.7\linewidth]{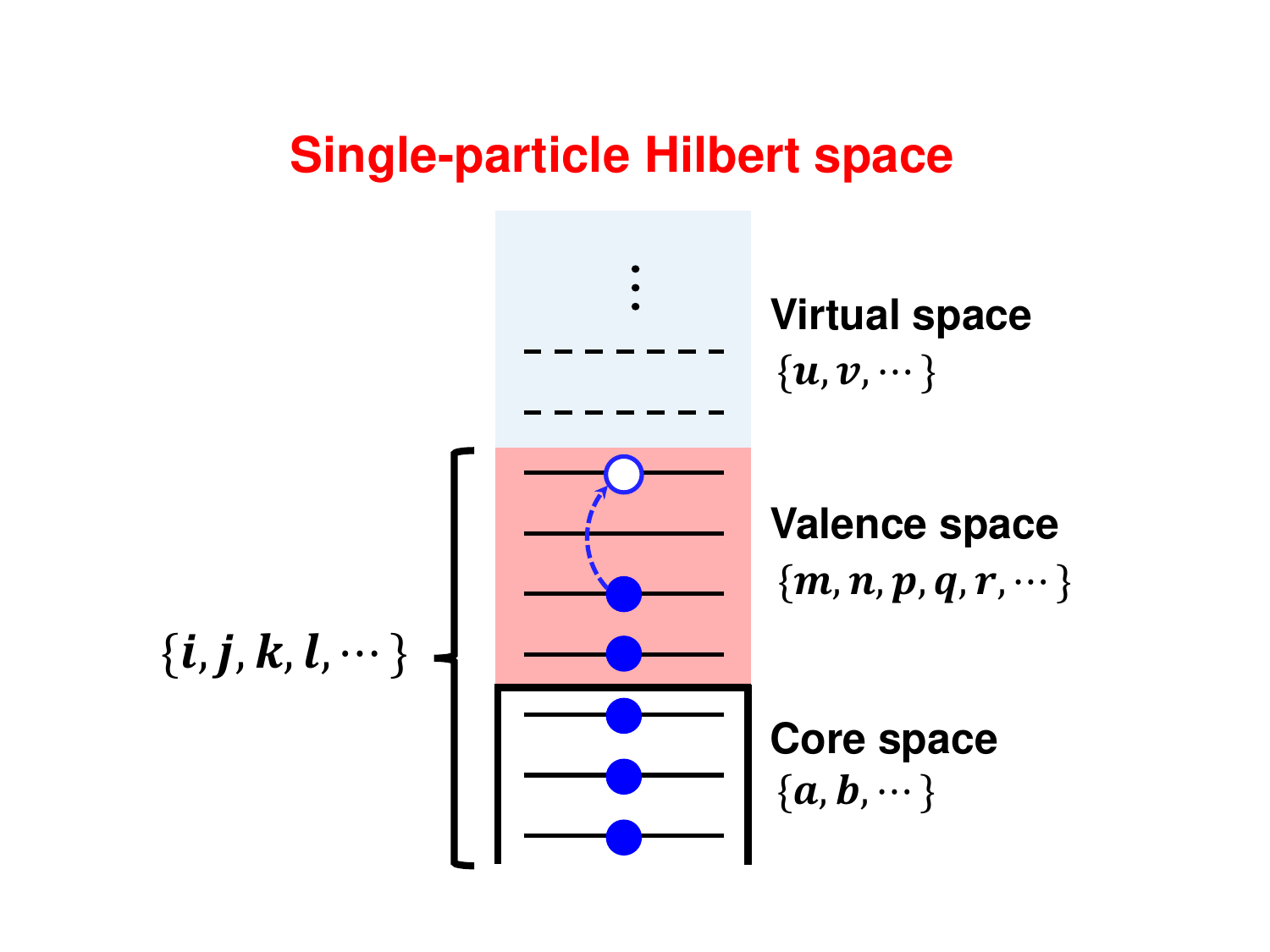}
   \caption{Illustration of the division of the single-particle Hilbert space. 
   We use distinct index sets to label different types of single-particle states: $\{a,b,\dots\}$ for core states, $\{m,n,p,q,r,\dots\}$ for valence states, and $\{u,v,\dots\}$ for virtual states. 
   Single-particle states in either the core or the valence space are labeled with indices $\{i,j,k,l,\dots\}$.}
   \label{Division-of-Hilbert-space}
 \end{figure*}

In principle, one can consider all possible configurations formed by distributing $N$ neutrons and $Z$ protons among the $N_{\text{spl}}$ single-particle states.
However, such a treatment is impractical for systems with dozens or even hundreds of nucleons, as the computational cost grows exponentially with the number of particles.
Therefore, we introduce a truncation by dividing the single-particle Hilbert space into the core space, the valence space, and the virtual space, as shown schematically in Fig. \ref{Division-of-Hilbert-space}.
By definition, single-particle states in the core space are occupied at all times, those in the valence space can have occupancies of either 0 or 1 for each configuration, and those in the virtual space remain unoccupied throughout the time evolution.
Importantly, the single-particle states in the core space are not frozen during the time evolution but evolve dynamically and self-consistently together with the states in the valence space.
To distinguish different spaces, we use distinct index sets to label different types of single-particle states: $\{a,b,\dots\}$ for core states, $\{m,n,p,q,r,\dots\}$ for valence states, and $\{u,v,\dots\}$ for virtual states. 
Single-particle states in either the core or the valence space are labeled with indices $\{i,j,k,l,\dots\}$.
Generic single-particle states, when no distinction is needed, are denoted by Greek letters $\{\alpha,\lambda,\xi,\zeta\dots\}$.

\subsection{The many-body model Hamiltonian}

The nuclear many-body Hamiltonian reads
\begin{equation}
\label{Hamiltonian}
\hat{H}=\hat{H}_{\text{DFT}}+\hat{H}_{\text{pair}},
\end{equation}
where $\hat{H}_{\text{DFT}}$ is derived from a universal density functional, and $\hat{H}_{\text{pair}}$ corresponds to the pairing interaction.

In this work, the relativistic density functional PC-PK1 \cite{ZhaoPW2010PRC} is employed, and the corresponding Hamiltonian $\hat{H}_{\text{DFT}}$ reads 
\begin{equation}\label{H}
  \hat{H}_{\text{DFT}} = \int \mathrm{d}^{3} r (\mathcal{H}^{\text{free}}+\mathcal{H}^{\text{4f}}+\mathcal{H}^{\text{der}}+\mathcal{H}^{\text{em}}+\mathcal{H}^{\text{hot}}). \\
\end{equation}
It includes the free-nucleon term $\mathcal{H}^{\text{free}}$,
\begin{equation}
\mathcal{H}^{\text{free}}=\bar{\psi}(-\mathrm{i} \boldsymbol{\gamma} \cdot \boldsymbol{\nabla}+m) \psi,
\end{equation}
the four-fermion point-coupling terms $\mathcal{H}^{\text{4f}}$,
\begin{equation}
\mathcal{H}^{\text{4f}}=\frac{1}{2} \alpha_{S}(\bar{\psi} \psi)(\bar{\psi} \psi)+\frac{1}{2} \alpha_{V}\left(\bar{\psi} \gamma_{\mu} \psi\right)\left(\bar{\psi} \gamma^{\mu} \psi\right)+\frac{1}{2} \alpha_{T V}\left(\bar{\psi} \vec{\tau} \gamma_{\mu} \psi\right)\left(\bar{\psi} \vec{\tau} \gamma^{\mu} \psi\right), 
\end{equation}
the gradient terms $\mathcal{H}^{\text{der}}$ that account for finite range effects,
\begin{equation}
\begin{split}
\mathcal{H}^{\text{der}}=&-\frac{1}{2} \delta_{S}\left[\partial_{0}(\bar{\psi} \psi) \partial^{0}(\bar{\psi} \psi)+\boldsymbol{\nabla}(\bar{\psi} \psi) \cdot \boldsymbol{\nabla}(\bar{\psi} \psi)\right] \\
      &-\frac{1}{2} \delta_{V}\left[\partial_{0}\left(\bar{\psi} \gamma_{\mu} \psi\right) \partial^{0}\left(\bar{\psi} \gamma^{\mu} \psi\right)+\boldsymbol{\nabla}\left(\bar{\psi} \gamma_{\mu} \psi\right) \cdot \boldsymbol{\nabla}\left(\bar{\psi} \gamma^{\mu} \psi\right)\right] \\
      &-\frac{1}{2} \delta_{T V}\left[\partial_{0}\left(\bar{\psi} \vec{\tau} \gamma_{\mu} \psi\right) \partial^{0}\left(\bar{\psi} \vec{\tau} \gamma^{\mu} \psi\right)+\boldsymbol{\nabla}\left(\bar{\psi} \vec{\tau} \gamma_{\mu} \psi\right) \cdot \boldsymbol{\nabla}\left(\bar{\psi} \vec{\tau} \gamma^{\mu} \psi\right)\right],\\
\end{split}
\end{equation}
the electromagnetic interaction terms $\mathcal{H}^{\text{em}}$,
\begin{equation}
\mathcal{H}^{\text{em}}=\frac{1}{2}A_{\mu}\Delta A^{\mu}+e \frac{1-\tau_{3}}{2} \bar{\psi} \gamma^{\mu} A_{\mu} \psi,
\end{equation}
and higher-order terms $\mathcal{H}^{\text{hot}}$ that lead to medium effects,
\begin{equation}
\mathcal{H}^{\text{hot}}=\frac{1}{3} \beta_{S}(\bar{\psi} \psi)^{3}+\frac{1}{4} \gamma_{S}(\bar{\psi} \psi)^{4}+\frac{1}{4} \gamma_{V}\left[\left(\bar{\psi} \gamma_{\mu} \psi\right)\left(\bar{\psi} \gamma^{\mu} \psi\right)\right]^{2}.
\end{equation}
Here, $m$ is the nucleon mass, $e$ is the charge unit for protons, and $\vec{\tau}$ is the isospin Pauli matrix.
$A_{\mu}$ is the four-vector potential of the electromagnetic field, and $\psi$ is the Dirac spinor field of the nucleon. In total, the PC-PK1 Lagrangian is determined by 9 adjustable coupling constants, $\alpha_S, \alpha_V, \alpha_{TV}, \delta_S, \delta_V, \delta_{TV}, \beta_{S}, \gamma_S$ and $\gamma_V$, that have been fine-tuned to properties of finite nuclei \cite{ZhaoPW2010PRC}. 
Within the no-retardation approximation \cite{Ring2001NPA}, the derivative terms are evaluated in their instantaneous static form. The  time-derivative parts of $\mathcal{H}^{\text{der}}$, such as $\partial_0(\bar\psi\psi)\partial^0(\bar\psi\psi)$, are therefore neglected.

Note that while $\mathcal{H}^{\text{free}}$, $\mathcal{H}^{\text{4f}}$, $\mathcal{H}^{\text{der}}$ and $\mathcal{H}^{\text{em}}$ include only one- and two-body interactions, $\mathcal{H}^{\text{hot}}$ contains three- and four-body interactions. 
Because the three- and four-body contributions lead to lengthy expressions, they are not displayed in the main derivation below. They are, however, included in the numerical implementation, and their explicit expressions are given in Appendix~B.

In the no-sea approximation, the nucleon field operator $\psi$ is expanded in terms of the positive-energy single-particle states:  $\psi = \sum_{\alpha,\tau}\phi_{\alpha,\tau}c_{\alpha,\tau}$. Correspondingly, the Hamiltonian $\hat{H}_{\text{DFT}}$ can be expressed in second-quantized form 
\begin{equation}
\hat{H}_{\text{DFT}}=\sum_{\tau}\sum_{\alpha\lambda}h_{\alpha\lambda}^{\tau}c_{\alpha,\tau}^{\dagger}c_{\lambda,\tau}+\frac{1}{2}\sum_{\tau\tau^{\prime}}\sum_{\alpha\lambda\xi\zeta}W_{\alpha\lambda\xi\zeta}^{\tau\tau^{\prime}}(c_{\alpha,\tau}^{\dagger}c_{\lambda,\tau})(c_{\xi,\tau^{\prime}}^{\dagger}c_{\zeta,\tau^{\prime}})+\int d^3r\,\frac{1}{2}A_\mu\Delta A^\mu,
\end{equation}
where the one-body and two-body Hamiltonian matrix elements read, respectively 
\begin{equation}
h_{\alpha\lambda}^{\tau}=\int d^3r \left[\bar{\phi}_{\alpha,\tau}(-\mathrm{i}\boldsymbol{\gamma}\cdot\boldsymbol{\nabla}+m)\phi_{\lambda,\tau}+e\frac{1-\tau_3}{2}\bar{\phi}_{\alpha,\tau}\gamma^{\mu}A_{\mu}\phi_{\lambda,\tau}\right],
\end{equation}
and
\begin{equation}
\begin{split}
W_{\alpha\lambda\xi\zeta}^{\tau\tau^{\prime}}=\int d^3r&\bigg\{\alpha_S(\bar{\phi}_{\alpha,\tau}\phi_{\lambda,\tau})(\bar{\phi}_{\xi,\tau^{\prime}}\phi_{\zeta,\tau^{\prime}})+\alpha_V(\bar{\phi}_{\alpha,\tau}\gamma_{\mu}\phi_{\lambda,\tau})(\bar{\phi}_{\xi,\tau^{\prime}}\gamma^{\mu}\phi_{\zeta,\tau^{\prime}})\\
&+\alpha_{TV}(\bar{\phi}_{\alpha,\tau}\vec{\tau}\gamma_{\mu}\phi_{\lambda,\tau})(\bar{\phi}_{\xi,\tau^{\prime}}\vec{\tau}\gamma^{\mu}\phi_{\zeta,\tau^{\prime}})-\delta_S\boldsymbol{\nabla}(\bar{\phi}_{\alpha,\tau}\phi_{\lambda,\tau})\cdot \boldsymbol{\nabla}(\bar{\phi}_{\xi,\tau^{\prime}}\phi_{\zeta,\tau^{\prime}}) \\
&-\delta_V\boldsymbol{\nabla}(\bar{\phi}_{\alpha,\tau}\gamma_{\mu}\phi_{\lambda,\tau})\cdot \boldsymbol{\nabla}(\bar{\phi}_{\xi,\tau^{\prime}}\gamma^{\mu}\phi_{\zeta,\tau^{\prime}})-\delta_{TV}\boldsymbol{\nabla}(\bar{\phi}_{\alpha,\tau}\vec{\tau}\gamma_{\mu}\phi_{\lambda,\tau})\cdot \boldsymbol{\nabla}(\bar{\phi}_{\xi,\tau^{\prime}}\vec{\tau}\gamma^{\mu}\phi_{\zeta,\tau^{\prime}})\bigg\}.
\end{split}
\end{equation}
For $\hat{H}_{\text{DFT}}$, the Hartree approximation is adopted in the following derivation, which means that Wick contractions between operators from different brackets are neglected. 

The long-range particle-hole interaction is taken into account within $\hat{H}_{\text{DFT}}$, while the short-range particle-particle interaction will be described by a phenomenological monopole pairing interaction:
\begin{equation}
\label{pairing-Hamiltonian}
\hat{H}_{\text{pair}}=-\sum_{\tau}G^{\tau}\left[\hat{P}_{\tau}^{\dagger}\hat{P}_{\tau}-\sum_{\mu>0}\hat{P}_{\mu,\tau}^{\dagger}\hat{P}_{\mu,\tau}\right],
\end{equation}
where,
\begin{equation}
\hat{P}_{\tau}^{\dagger}=\sum_{r,s>0}d_{r,s,\tau}^{\dagger}d_{r,\bar{s},\tau}^{\dagger},\qquad \hat{P}_{\mu,\tau}^{\dagger}=e_{\mu,\tau}^{\dagger}e_{\bar{\mu},\tau}^{\dagger}.
\end{equation}
Here, $G^{\text{n}(\text{p})}$ is the effective pairing strength for neutrons (protons). The first term represents the standard monopole pairing interaction, with $d_{r,s,\text{n}(\text{p})}^{\dagger}$ creating a neutron(proton) at lattice point $r$ with spin $s$. Here, $\bar{s}$ denotes the time-reversal spin state.
The second term is introduced to remove the self-scattering contribution of nucleon pairs, which would otherwise give an
unphysical contribution to the pairing energy. 
In this term, $e_{\mu,\tau}^{\dagger}$ represents the creation operator of the single-particle state at the initial time $t_0$, which is labeled as $|\varphi_{\mu,\tau}\rangle$, and $\bar{\mu}$ labels the time-reversal state of $\mu$.
Note that the pairing Hamiltonian \eqref{pairing-Hamiltonian} is essentially the same as the standard one in RDFT-SLAP \cite{MengJ2006FPC}.
The only difference lies in the basis in which it is expanded.

Using the unitary transformations $d_{r,s,\tau}^{\dagger}=\sum_{\alpha}\langle\phi_{\alpha,\tau}|r,s,\tau\rangle c_{\alpha,\tau}^{\dagger}$ and $e_{\mu,\tau}^{\dagger}=\sum_{\alpha}\langle\phi_{\alpha,\tau}|\varphi_{\mu,\tau}\rangle c_{\alpha,\tau}^{\dagger}$, $\hat{H}_{\text{pair}}$ can be rewritten in second-quantized form as follows
\begin{equation}
\hat{H}_{\text{pair}}=\sum_{\tau}\sum_{\alpha\lambda\xi\zeta}Q_{\alpha\lambda\xi\zeta}^{\tau}c_{\alpha,\tau}^{\dagger}c_{\lambda,\tau}^{\dagger}c_{\xi,\tau}c_{\zeta,\tau},
\end{equation}
with  
\begin{equation}
\begin{split}
Q_{\alpha\lambda\xi\zeta}^{\tau}=&-\sum_{r_1,s_1>0}\sum_{r_2,s_2>0}G^{\tau}\langle\phi_{\alpha,\tau}|r_1,s_1,\tau\rangle\langle\phi_{\lambda,\tau}|r_1,\bar{s}_1,\tau\rangle\langle r_2,\bar{s}_2,\tau|\phi_{\xi,\tau}\rangle\langle r_2,s_2,\tau|\phi_{\zeta,\tau}\rangle\\
&+\sum_{\mu>0}G^{\tau}\langle\phi_{\alpha,\tau}|\varphi_{\mu,\tau}\rangle\langle\phi_{\lambda,\tau}|\varphi_{\bar{\mu},\tau}\rangle\langle \varphi_{\bar{\mu},\tau}|\phi_{\xi,\tau}\rangle\langle \varphi_{\mu,\tau}|\phi_{\zeta,\tau}\rangle.\\
\end{split}
\end{equation}

\subsection{Variational equations of the action functional}

To model the dynamics of a nuclear system described by the correlated wave function \eqref{wave-function}, we need to derive the set of equations of motion for the expansion coefficients and single-particle states.
To this end, we follow the Dirac-Frenkel time-dependent variational principle \cite{Dirac1930Cambridge,Frenkel1934Wave}. 
First, the action functional is defined: 
\begin{equation}
S\left[\left\{C_I\right\}, \left\{\phi_{i,\tau}\right\}, \left\{\mu_{ij}^{\tau}\right\}\right]=\int_{t=t_0}^{t_1}dt\left[ \left\langle\Psi\left|\left(\hat{H}-i\hbar\partial_t\right)\right|\Psi\right\rangle-\sum_{\tau}\sum_{ij}\mu_{ij}^{\tau}\left(\left\langle\phi_{i,\tau}|\phi_{j,\tau}\right\rangle-\delta_{ij}\right) \right].
\end{equation}
Here, $\mu_{ij}^{\tau}$ is the Lagrange multiplier that ensures orthonormality of the core and valence states during time evolution. The variation of the action functional reads
\begin{equation}
\label{deltaS}
\begin{split}
\delta S=&\int_{t=t_0}^{t_1}dt \Bigg[\left\langle\delta\Psi\left|\left(\hat{H}-i\hbar\partial_t\right)\right|\Psi\right\rangle+\left\langle\Psi\left|\left(\hat{H}+i\hbar\overleftarrow{\partial_t}\right)\right|\delta\Psi\right\rangle\\
&~~\qquad\quad-\sum_{\tau}\sum_{ij}\mu_{ij}^{\tau}\left(\langle\delta\phi_{i,\tau}|\phi_{j,\tau}\rangle+\langle\phi_{i,\tau}|\delta\phi_{j,\tau}\rangle\right)-\sum_{\tau}\sum_{ij}\delta\mu_{ij}^{\tau}\left(\langle\phi_{i,\tau}|\phi_{j,\tau}\rangle-\delta_{ij}\right) \Bigg],\\
\end{split}
\end{equation}
where a time-derivative operator with the arrow pointing to the left denotes its action on the bra-vector.
By taking into account the completeness and orthonormality of single-particle states, the variation and time-derivative of a single-particle state $|\phi_{\lambda,\tau}\rangle$ can be written: $|\delta{\phi}_{\lambda,\tau}\rangle=\sum_{\alpha}|\phi_{\alpha,\tau}\rangle\langle\phi_{\alpha,\tau}|\delta{\phi}_{\lambda,\tau}\rangle$ and $|\dot{\phi}_{\lambda,\tau}\rangle=\sum_{\alpha}|\phi_{\alpha,\tau}\rangle\langle\phi_{\alpha,\tau}|\dot{\phi}_{\lambda,\tau}\rangle$, respectively. 
Then the variation and the time-derivative of the total wave function \eqref{wave-function} are given by \cite{Miyagi2013PRA} 
\begin{subequations}
\begin{equation}
\label{delta-Psi}
|\delta\Psi\rangle=\sum_{I}\delta C_I|\text{MPC}_I\rangle+\sum_{\tau}\sum_{\alpha\lambda}c_{\alpha,\tau}^{\dagger}c_{\lambda,\tau}|\Psi\rangle\langle\phi_{\alpha,\tau}|\delta \phi_{\lambda,\tau}\rangle,
\end{equation}
\begin{equation}
\label{partial-Psi}
\partial_t|\Psi\rangle=\sum_{I}\dot{C}_I|\text{MPC}_I\rangle+\sum_{\tau}\sum_{\alpha\lambda}c_{\alpha,\tau}^{\dagger}c_{\lambda,\tau}|\Psi\rangle\langle\phi_{\alpha,\tau}|\dot{\phi}_{\lambda,\tau}\rangle.
\end{equation}
\end{subequations}
The anti-Hermitian matrix $\eta_{\alpha\lambda}^{\tau}\equiv\langle\phi_{\alpha,\tau}|\dot{\phi}_{\lambda,\tau}\rangle$ will play an important role in the following derivation of the equations of motion for single-particle states.

The variation of the action functional with respect to small variations of $\langle\phi_{i,\tau}|$ and $C_J^*$,
\begin{subequations}
\begin{equation}
\label{variation-phi}
\frac{\delta S}{\langle\delta\phi_{i,\tau}|}=0,
\end{equation}
\begin{equation}
\label{variation-C}
\frac{\delta S}{\delta C_J^*}=0,
\end{equation}
\end{subequations}
leads to the equations of motion for single-particle states and expansion coefficients, respectively. 
Correspondingly, the stationary conditions $\delta S/|\delta\phi_{i,\tau}\rangle=\delta S/\delta C_J=0$ result in a set of equations that are Hermitian conjugates of the set obtained from Eqs. \eqref{variation-phi} and \eqref{variation-C}.

\subsection{Equations of motion for single-particle states}

After inserting Eqs. \eqref{delta-Psi} and \eqref{partial-Psi} and their Hermitian conjugates into Eq. \eqref{deltaS}, from the stationary condition \eqref{variation-phi} one obtains the equations of motion for single-particle states   
\begin{equation}
\label{EOM-VS}
\sum_{\alpha}|\phi_{\alpha,\tau}\rangle\langle\Psi|c_{i,\tau}^{\dagger}c_{\alpha,\tau}\left[-\mathrm{i}\hbar\sum_I\dot{C}_I|\text{MPC}_I\rangle+\left(\hat{H}-\mathrm{i}\hbar\sum_{\tau^{\prime}}\sum_{\xi\zeta}\eta_{\xi\zeta}^{\tau^{\prime}} c_{\xi,\tau^{\prime}}^{\dagger}c_{\zeta,\tau^{\prime}}\right)|\Psi\rangle\right]-\sum_j\mu_{ij}^{\tau}|\phi_{j,\tau}\rangle=0.
\end{equation}
Multiplying Eq. \eqref{EOM-VS} from the left by a virtual state $\langle \phi_{u,\tau}|$, and using the orthogonality of single-particle states, this expression is simplified to 
\begin{equation}
\label{EOM-VS-2}
\left\langle\Psi\left|c_{i,\tau}^{\dagger}c_{u,\tau}\left(\hat{H}-\mathrm{i}\hbar\sum_{\tau^{\prime}}\sum_{\xi\zeta}\eta_{\xi\zeta}^{\tau^{\prime}}c_{\xi,\tau^{\prime}}^{\dagger}c_{\zeta,\tau^{\prime}}\right)\right|\Psi\right\rangle=0.
\end{equation}
Substituting the Hamiltonian \eqref{Hamiltonian} into Eq. \eqref{EOM-VS-2} and performing some algebra with the help of Wick’s theorem \cite{Wick1950PR}, the equations of motion for single-particle states read
\begin{equation}
\label{EOM-sp-VS}
\sum_ji\hbar\eta_{uj}^{\tau}\rho_{ij}^{\tau}=\sum_jh_{uj}^{\tau}\rho_{ij}^{\tau}+\sum_{\tau^{\prime}}\sum_{jkl}W_{ujkl}^{\tau\tau^{\prime}}\rho_{ijkl}^{\tau\tau^{\prime}}+\sum_{jkl}\tilde{Q}_{ujkl}^{\tau}\kappa_{ijkl}^{\tau},
\end{equation}
with $\tilde{Q}_{ujkl}^{\tau}={Q}_{ujkl}^{\tau}-{Q}_{jukl}^{\tau}$. Here, $\rho_{ij}^{\tau}$ denotes the one-body density matrix. The matrices $\rho_{ijkl}^{\tau\tau^{\prime}}$ and $\kappa_{ijkl}^{\tau}$ represent the two-body density matrices for the particle-hole and particle-particle channels, respectively. 
They are defined as follows:
\begin{subequations}
\begin{align}
&\rho_{ij}^{\tau}=\langle\Psi|c_{i,\tau}^{\dagger}c_{j,\tau}|\Psi\rangle,\\
&\rho_{ijkl}^{\tau\tau^{\prime}}=\langle\Psi|(c_{i,\tau}^{\dagger}c_{j,\tau})(c_{k,\tau^{\prime}}^{\dagger}c_{l,\tau^{\prime}})|\Psi\rangle,\\
&\kappa_{ijkl}^{\tau}=\langle\Psi|c_{i,\tau}^{\dagger}c_{j,\tau}^{\dagger}c_{k,\tau}c_{l,\tau}|\Psi\rangle.
\end{align}
\end{subequations}
The explicit expressions for these density matrices are given in Appendix A.

When the core states are considered, the one-body density matrix and two-body density matrix for the particle-hole channel read (see Appendix A):
\begin{equation}
\rho_{ab}^{\tau}=\delta_{ab},\qquad \rho_{aabb}^{\tau\tau^{\prime}}=1,\qquad \rho_{aamn}^{\tau\tau^{\prime}}=\rho_{mn}^{\tau^{\prime}},\qquad \rho_{mnaa}^{\tau\tau^{\prime}}=\rho_{mn}^{\tau}. 
\end{equation}
Therefore, the equations \eqref{EOM-sp-VS} can be simplified and divided into two sets: 
\begin{subequations}
\begin{equation}
\label{eta-ua-ur-1}
 \mathrm{i}\hbar\eta_{ua}^{\tau}=h_{ua}^{\tau}+\sum_{\tau^{\prime}}\sum_b W_{uabb}^{\tau\tau^{\prime}}+\sum_{\tau^{\prime}}\sum_{mn}\rho_{mn}^{\tau^{\prime}}W_{uamn}^{\tau\tau^{\prime}}+\sum_{jkl}\kappa_{ajkl}^{\tau}\tilde{Q}_{ujkl}^{\tau},\\
 \end{equation}
 \begin{equation}
 \label{eta-ua-ur-2}
 \mathrm{i}\hbar\eta_{ur}^{\tau}=h_{ur}^{\tau}+\sum_{\tau^{\prime}}\sum_b W_{urbb}^{\tau\tau^{\prime}}+\sum_{\tau^{\prime}}\sum_{mnpq}(\rho^{\tau}_{rm})^{-1}\rho_{mnpq}^{\tau\tau^{\prime}}W_{unpq}^{\tau\tau^{\prime}}+\sum_{mjkl}(\rho^{\tau}_{rm})^{-1}\kappa_{mjkl}^{\tau}\tilde{Q}_{ujkl}^{\tau},
\end{equation}
\end{subequations}
where $(\rho^{\tau}_{rm})^{-1}$ denotes the inverse of the one-body density matrix $\rho^{\tau}_{rm}$, that is, $\sum_m (\rho^\tau)^{-1}_{rm}\rho^\tau_{mn}=\delta_{rn}$.

To circumvent explicit numerical treatment of virtual states, we introduce a projector on the space spanned by virtual states $\hat{P}_{\tau}=\sum_u|\phi_{u,\tau}\rangle\langle\phi_{u,\tau}|=1-\sum_i|\phi_{i,\tau}\rangle\langle\phi_{i,\tau}|$, and express Eqs. \eqref{eta-ua-ur-1} and \eqref{eta-ua-ur-2} in the following form
\begin{subequations}
\begin{equation}
\label{Pphi-a}
\mathrm{i}\hbar\hat{P}_{\tau}|\dot{\phi}_{a,\tau}\rangle=\hat{P}_{\tau}\left\{\left(\hat{h}+\hat{W}_{\text{core}}\right)|\phi_{a,\tau}\rangle+\sum_{\tau^{\prime}}\sum_{mn}\rho_{mn}^{\tau^{\prime}}\hat{W}_{mn}^{\tau\tau^{\prime}}|\phi_{a,\tau}\rangle+\sum_{jkl}\kappa_{ajkl}^{\tau}\langle\phi_{j,\tau}|\tilde{Q}_{kl}^{\tau}\right\},\\
\end{equation}
\begin{equation}
\label{Pphi-r}
\begin{aligned}
\mathrm{i}\hbar\hat{P}_{\tau}|\dot{\phi}_{r,\tau}\rangle=\hat{P}_{\tau}&\left\{\left(\hat{h}+\hat{W}_{\text{core}}\right)|\phi_{r,\tau}\rangle+\sum_{\tau^{\prime}}\sum_{mnpq}(\rho^{\tau}_{rm})^{-1}\rho_{mnpq}^{\tau\tau^{\prime}}\hat{W}_{pq}^{\tau\tau^{\prime}}|\phi_{n,\tau}\rangle\right.\\
&\qquad\qquad\qquad\qquad\qquad\qquad\qquad\qquad\left.+\sum_{mjkl}(\rho_{rm}^{\tau})^{-1}\kappa_{mjkl}^{\tau}\langle\phi_{j,\tau}|\tilde{Q}_{kl}^{\tau}\right\}.\\
\end{aligned}
\end{equation}
\end{subequations}
For the particle-hole channel, the one-body operator $\hat{h}$, and the potentials resulting from the core states $\hat{W}_{\text{core}}$ and the valence states $\hat{W}_{pq}^{\tau\tau^{\prime}}$ are defined: 
\begin{subequations}
\begin{align}
&\hat{h}=\boldsymbol{\alpha}\cdot\hat{\boldsymbol{p}}+\beta m+e\frac{1-\tau_3}{2}\gamma_0\gamma^{\mu}A_{\mu},\\
&\hat{W}_{\text{core}}=\beta S_{\text{core}}+V^0_{\text{core}}-\boldsymbol{\alpha}\cdot\boldsymbol{V}_{\text{core}},\\
&\hat{W}_{pq}^{\tau\tau^{\prime}}=\beta S_{pq}^{\tau^{\prime}}+(V^0)_{pq}^{\tau\tau^{\prime}}-\boldsymbol{\alpha}\cdot (\boldsymbol{V})_{pq}^{\tau\tau^{\prime}},
\end{align}
\end{subequations}
where the scalar potentials $S_{\text{core}}$ and $S_{pq}^{\tau^{\prime}}$, and the four-vector potentials $V^{\mu}_{\text{core}}$ and  $(V^{\mu})_{pq}^{\tau\tau^{\prime}}$, are given by
\begin{subequations}
\begin{align}
&S_{\text{core}}=\alpha_S(\rho_S)_{\text{core}}+\delta_S\Delta(\rho_S)_{\text{core}},\\
&S_{pq}^{\tau^{\prime}}=\alpha_S(\rho_S)_{pq}^{\tau^{\prime}}+\delta_S\Delta(\rho_S)_{pq}^{\tau^{\prime}},\\
&V^{\mu}_{\text{core}}=\alpha_V(j^{\mu})_{\text{core}}+\delta_V\Delta (j^{\mu})_{\text{core}}+\tau_3\alpha_{TV}(j_{TV}^{\mu})_{\text{core}}+\tau_3\delta_{TV}(\Delta j_{TV}^{\mu})_{\text{core}},\\
&(V^{\mu})_{pq}^{\tau\tau^{\prime}}=\alpha_V(j^{\mu})_{pq}^{\tau^{\prime}}+\delta_V\Delta(j^{\mu})_{pq}^{\tau^{\prime}}+\tau_3\alpha_{TV}(j_{TV}^{\mu})_{pq}^{\tau^{\prime}}+\tau_3\delta_{TV}\Delta(j_{TV}^{\mu})_{pq}^{\tau^{\prime}}.
\end{align}
\end{subequations}
The densities and currents are defined in terms of the core and valence single-particle states as follows:
\begin{subequations}
\begin{align}
&(\rho_S)_{\text{core}}(\boldsymbol{r},t)=\sum_{\tau}\sum_a\bar{\phi}_{a,\tau}(\boldsymbol{r},t)\phi_{a,\tau}(\boldsymbol{r},t);\qquad (\rho_S)_{pq}^{\tau}(\boldsymbol{r},t)=\bar{\phi}_{p,\tau}(\boldsymbol{r},t)\phi_{q,\tau}(\boldsymbol{r},t);\\
&(j^{\mu})_{\text{core}}(\boldsymbol{r},t)=\sum_{\tau}\sum_a\bar{\phi}_{a,\tau}(\boldsymbol{r},t)\gamma^{\mu}\phi_{a,\tau}(\boldsymbol{r},t);\qquad (j^{\mu})_{pq}^{\tau}(\boldsymbol{r},t)=\bar{\phi}_{p,\tau}(\boldsymbol{r},t)\gamma^{\mu}\phi_{q,\tau}(\boldsymbol{r},t);\\
&(j_{TV}^{\mu})_{\text{core}}(\boldsymbol{r},t)=\sum_{\tau}\sum_a\bar{\phi}_{a,\tau}(\boldsymbol{r},t)\gamma^{\mu}\tau_3\phi_{a,\tau}(\boldsymbol{r},t);\qquad (j_{TV}^{\mu})_{pq}^{\tau}(\boldsymbol{r},t)=\bar{\phi}_{p,\tau}(\boldsymbol{r},t)\gamma^{\mu}\tau_3\phi_{q,\tau}(\boldsymbol{r},t).
\end{align}
\end{subequations}
For the particle-particle channel, we introduce $\tilde{Q}_{kl}^{\tau}$: 
\begin{equation}
\begin{split}
\tilde{Q}_{kl}^{\tau}=&-\sum_{r_1,s_1>0}\sum_{r_2,s_2>0}G^{\tau}|r_1,\bar{s}_1,\tau\rangle|r_1,s_1,\tau\rangle\langle r_2,\bar{s}_2,\tau|\phi_{k,\tau}\rangle\langle r_2,s_2,\tau|\phi_{l,\tau}\rangle\\
&+\sum_{r_1,s_1>0}\sum_{r_2,s_2>0}G^{\tau}|r_1,{s}_1,\tau\rangle|r_1,\bar{s}_1,\tau\rangle\langle r_2,\bar{s}_2,\tau|\phi_{k,\tau}\rangle\langle r_2,s_2,\tau|\phi_{l,\tau}\rangle\\
&+\sum_{\mu>0}G^{\tau}|\varphi_{\bar{\mu},\tau}\rangle|\varphi_{\mu,\tau}\rangle\langle\varphi_{\bar{\mu},\tau}|\phi_{k,\tau}\rangle\langle\varphi_{\mu,\tau}|\phi_{l,\tau}\rangle-\sum_{\mu>0}G^{\tau}|\varphi_{{\mu},\tau}\rangle|\varphi_{\bar{\mu},\tau}\rangle\langle\varphi_{\bar{\mu},\tau}|\phi_{k,\tau}\rangle\langle\varphi_{\mu,\tau}|\phi_{l,\tau}\rangle.\\
\end{split}
\end{equation}
Here, the first ket in $\tilde Q^\tau_{kl}$ is contracted with $\langle\phi_{j,\tau}|$, whereas the second one is contracted with $\langle\phi_{u,\tau}|$ in Eqs. \eqref{Pphi-a} and \eqref{Pphi-r}.

In Eqs. \eqref{Pphi-a} and \eqref{Pphi-r}, the projector $\hat{P}_{\tau}$ appears both on the left and right sides, resulting in cumbersome coupled systems of integro-differential nonlinear equations. 
The operator on the left-hand side can be eliminated if $\eta_{ij}^{\tau}=0$.
As shown in Ref. \cite{Alon2008PRA}, since the nuclear wave function \eqref{wave-function} is invariant under unitary transformations of single-particle states in the same space, one can always find a unitary matrix that yields a set of states satisfying $\eta_{ab}^{\tau}=\eta_{mn}^{\tau}=0$. 
The values of $\eta_{am}^{\tau}$ and $\eta_{ma}^{\tau}$ are computed (see Appendix C), and turn out to be very small in this case; they are therefore set to zero.
Finally, our working equations of motion for single-particle states read
\begin{subequations}
\begin{equation}
\label{phi-a}
\mathrm{i}\hbar|\dot{\phi}_{a,\tau}\rangle=\hat{P}_{\tau}\left\{\left(\hat{h}+\hat{W}_{\text{core}}\right)|\phi_{a,\tau}\rangle+\sum_{\tau^{\prime}}\sum_{mn}\rho_{mn}^{\tau^{\prime}}\hat{W}_{mn}^{\tau\tau^{\prime}}|\phi_{a,\tau}\rangle+\sum_{jkl}\kappa_{ajkl}^{\tau}\langle\phi_{j,\tau}|\tilde{Q}_{kl}^{\tau}\right\},\\
\end{equation}
\begin{equation}
\label{phi-r}
\begin{aligned}
\mathrm{i}\hbar|\dot{\phi}_{r,\tau}\rangle=\hat{P}_{\tau}&\left\{\left(\hat{h}+\hat{W}_{\text{core}}\right)|\phi_{r,\tau}\rangle+\sum_{\tau^{\prime}}\sum_{mnpq}(\rho^{\tau}_{rm})^{-1}\rho_{mnpq}^{\tau\tau^{\prime}}\hat{W}_{pq}^{\tau\tau^{\prime}}|\phi_{n,\tau}\rangle\right.\\
&\qquad\qquad\qquad\qquad\qquad\qquad\qquad\qquad\left.+\sum_{mjkl}(\rho_{rm}^{\tau})^{-1}\kappa_{mjkl}^{\tau}\langle\phi_{j,\tau}|\tilde{Q}_{kl}^{\tau}\right\}.\\
\end{aligned}
\end{equation}
\end{subequations} 
Note that TDDFT is a special case of the CI-TDDFT method. 
When no states are occupied in the valence space and pairing correlations are neglected, equations \eqref{phi-a} and \eqref{phi-r} reduce to the standard TDDFT equation \cite{RenZX2020PRC}.

\subsection{Equations of motion for the expansion coefficients}

The second stationary condition Eq. \eqref{variation-C} yields the equations of motion for the expansion coefficients,
\begin{equation}
\label{EOM-C}
\mathrm{i}\hbar\dot{C}_J=\sum_I(\mathcal{H}_{JI}-\mathcal{H}_{JI}^{\text{MF}})C_I,
\end{equation}
where $\mathcal{H}_{JI}$ is the Hamiltonian kernel, and $\mathcal{H}_{JI}^{\text{MF}}$ is the time-derivative kernel.
The Hamiltonian kernel reads
\begin{equation}
\begin{split}
\mathcal{H}_{JI}=&\langle\text{MPC}_J|\hat{H}|\text{MPC}_I\rangle\\
=&\sum_{\tau}\sum_{ij}h_{ij}^{\tau}\langle\text{MPC}_J|c_{i,\tau}^{\dagger}c_{j,\tau}|\text{MPC}_I\rangle+\frac{1}{2}\sum_{\tau\tau^{\prime}}\sum_{ijkl}W_{ijkl}^{\tau\tau^{\prime}}\langle\text{MPC}_J|(c_{i,\tau}^{\dagger}c_{j,\tau})(c_{k,\tau^{\prime}}^{\dagger}c_{l,\tau^{\prime}})|\text{MPC}_I\rangle\\
&+\sum_{\tau}\sum_{ijkl}Q_{ijkl}^{\tau}\langle\text{MPC}_J|c_{i,\tau}^{\dagger}c_{j,\tau}^{\dagger}c_{k,\tau}c_{l,\tau}|\text{MPC}_I\rangle+\delta_{JI}\int d^3r\,\frac{1}{2}A_\mu\Delta A^\mu,
\end{split}
\end{equation}
and the time-derivative kernel reads
\begin{equation}
\mathcal{H}_{JI}^{\text{MF}}=\sum_{\tau}\sum_{ij}\langle\text{MPC}_J|\mathrm{i}\hbar\eta_{ij}^{\tau}c_{i,\tau}^{\dagger}c_{j,\tau}|\text{MPC}_I\rangle.
\end{equation}
The time-derivative kernel $\mathcal{H}_{JI}^{\text{MF}}$ vanishes at all times for $\eta_{ij}^{\tau}=0$.

\subsection{Calculation of observables}

The expectation value of an observable $\hat{\mathcal{O}}$ in the correlated state \eqref{wave-function} at time $t$ reads
\begin{equation}
\label{observable-expectation}
\langle\Psi(t)|\hat{\mathcal{O}}|\Psi(t)\rangle=\sum_{JI}C_J^*(t)C_I(t)\langle\text{MPC}_J(t)|\hat{\mathcal{O}}|\text{MPC}_I(t)\rangle.
\end{equation}
This expression will be used, in this work for instance, to evaluate the time evolution of the energy and radius of a nuclear system. 

We close this section by emphasizing that the total energy, $E=\langle\Psi(t)|\hat{H}|\Psi(t)\rangle$, is conserved as a consequence of the fully variational evolution of both the expansion coefficients and the single-particle states.
Moreover, because of the choice of the wave function \eqref{wave-function}, the nucleon number is conserved even with the inclusion of pairing correlations. 
In the numerical implementation, the total energy and particle number conservation are preserved up to the accuracy of the adopted numerical scheme, as tested explicitly in Sec.~\ref{Numerical-test}.

\section{Numerical details}
\label{Numerical-details}

As a first application, we perform an illustrative study of excitation energies and widths of giant monopole resonances in $^{58}$Ni and $^{60}$Ni.
All calculations are carried out on a lattice in 3D coordinate space \cite{RenZX2017PRC,RenZX2019SCP,RenZX2020NPA,LiB2020PRC,XuFF2024PRC}, with a mesh spacing of 1.0 fm for all directions, and a lattice size of $L_x\times L_y\times L_z=24\times 24\times 24~\text{fm}^3$. 
For the particle-hole channel, the point-coupling relativistic density functional PC-PK1 \cite{ZhaoPW2010PRC} is adopted.
In the particle-particle channel, the effective pairing strengths $G^{\text{n}}$ and $G^{\text{p}}$ are determined by standard RDFT-SLAP calculations \cite{MengJ2006FPC,WangYP2023PLB,XuFF2024PRL}, with the strengths adjusted to reproduce the experimental odd-even mass differences within a specified MPC space.  
$^{58}$Ni and $^{60}$Ni are proton closed-shell nuclei with $Z=28$. 
Therefore, in this work, all single-proton states are treated as core states, and proton pairing correlations vanish ($G^{\text{p}}=0.0$ MeV).
For neutron configuration spaces, two kinds of MPC spaces, consisting of $N_{\text{MPC}} = 6$ and $15$ configurations, will be considered for both $^{58}$Ni and $^{60}$Ni.
The explicit configurations included in these spaces will be described in detail in Sec.~\ref{Results-and-discussions}.
For different MPC spaces, Table \ref{tab2} lists the adopted neutron pairing strengths, the odd-even mass differences calculated using the three-point formula, and the calculated binding energies and matter radii, together with the experimental results for comparison. The binding energies and matter radii of $^{58}$Ni and $^{60}$Ni calculated with RDFT-SLAP are in good agreement with the data.

\begin{table}
\centering
\caption{Adopted neutron pairing strengths (in MeV), odd-even mass differences calculated using the three-point formula (in MeV), and calculated binding energies (in MeV) and matter radii (in fm) for different MPC spaces, in comparison with the experimental data \cite{HuangWJ2021CPC,Wang2021CPC,LiLY2025ADNDT}.~\\}
\label{tab2}
  \begin{tabular}{ccccccccc}
     \toprule
     ~~Nucleus~~ & ~~$N_{\text{MPC}}$~~ & ~~$G^{\text{n}}$~~ & ~~$\Delta_{\text{cal.}}^{\text{n}}$~~ & ~~$\Delta_{\text{exp.}}^{\text{n}}$~~ & ~~$B_{\text{cal.}}$~~ & ~~$B_{\text{exp.}}$~~ & ~~$R_{\text{cal.}}$~~ & ~~$R_{\text{exp.}}$~~  \\
     \midrule
     $^{58}$Ni & 6 & 2.20 & 1.60 & 1.61 & 504.52 & 506.46 & 3.666 & 3.677(8) \\
     $^{58}$Ni  & 15 & 1.45 & 1.61 & 1.61 & 504.72 & 506.46 & 3.662 & 3.677(8)  \\
       $^{60}$Ni & 6 & 2.40 & 1.74 & 1.78 & 524.04 & 526.85 & 3.746 & 3.682(16)  \\
      $^{60}$Ni & 15 & 1.65 & 1.76 & 1.78 & 524.28 & 526.85 & 3.738 & 3.682(16)   \\
     \bottomrule
\end{tabular}
\end{table}
 
In the numerical implementation of the equations of motion for the expansion coefficients \eqref{EOM-C} and single-particle states \eqref{phi-a} and \eqref{phi-r}, the evolution time is divided into a series of short steps $\Delta t$.
Since the density matrices, potentials, and Hamiltonian kernels generally vary slowly with time, they are treated as constant over each time interval $[t, t+\Delta t]$.
This procedure is known as the constant mean field (CMF) approximation in quantum chemistry \cite{Beck1997ZPD}.
Over the CMF time step $\Delta t$, the equations of motion for the expansion coefficients and single-particle states are integrated separately.
More specifically, the equations for the expansion coefficients are linear and can be integrated only once within $[t, t+\Delta t]$, while those for the single-particle states are nonlinear and necessitate a finer time step $\Delta t'$.
This smaller time step is denoted as $\Delta t' = \Delta t / n$, with $n = 1, 2, 3, \dots$.
The values of $\Delta t$ and $\Delta t'$ should be chosen carefully to ensure the conservation of energy and particle number during time evolution.
Section \ref{Numerical-test} presents a detailed numerical check for the time steps $\Delta t$ and $\Delta t'$.
Unless otherwise specified, the values $\Delta t = 0.05$ fm/$c$ and $\Delta t' = \Delta t / 8$ are adopted in the following calculations.

For each time interval $[t, t+\Delta t]$, a predictor-corrector scheme is employed.
In this scheme, the expansion coefficients $C_J(t+\Delta t)$ and single-particle states $\phi_{i,\tau}(t+\Delta t)$ are determined using a two-step recipe.
In the first predictor step, using the density matrices, potentials, and Hamiltonian kernels at time $t$, the predicted expansion coefficients $\tilde{C}_J(t+\Delta t)$ and single-particle states $\tilde{\phi}_{i,\tau}(t+\Delta t)$ are obtained via a second-order Taylor expansion \cite{Barton1971CJ}.
These predicted values $\tilde{C}_J(t+\Delta t)$ and $\tilde{\phi}_{i,\tau}(t+\Delta t)$ are then used to provide estimates for the density matrices, potentials, and Hamiltonian kernels at $t+\Delta t$.
In the second, corrector step, the wave functions are propagated using the average values of $\mathcal{H}_{JI}$, $\hat{W}_{\text{core}}$, $\rho_{mn}^{\tau^{\prime}}\hat{W}_{mn}^{\tau\tau^{\prime}}$ and $(\rho_{rm}^{\tau})^{-1}\rho_{mnpq}^{\tau\tau^{\prime}}\hat{W}_{pq}^{\tau\tau^{\prime}}$ at $t$ and $t+\Delta t$. 
In this step, a more precise fourth-order Taylor expansion method is adopted.

Three additional modifications are introduced for numerical reasons. During time evolution, accumulated numerical errors may cause the single-particle states to become non-orthonormal. In such cases, the projection operator $\hat{P}_{\tau}=1-\sum_i|\phi_{i,\tau}\rangle\langle\phi_{i,\tau}|$ ceases to be a projector, and even an exact solution of Eqs. \eqref{phi-a} and \eqref{phi-r} will then further destroy orthonormality.
As proposed in Ref. \cite{Beck1997ZPD}, a remedy is to define an optimized projection operator as 
\begin{equation}
\label{opt-projector}
\hat{P}_{\tau}^{\text{opt}}=1-\sum_{ij}|\phi_{i,\tau}\rangle(\mathcal{R}_{ij}^{\tau})^{-1}\langle\phi_{j,\tau}|,\qquad \mathcal{R}_{ij}^{\tau}=\langle\phi_{i,\tau}|\phi_{j,\tau}\rangle.
\end{equation}
This optimized operator remains a projector as long as the single-particle states are linearly independent.

The second modification concerns the inverse of the one-body density matrix.
During time evolution, the one-body density matrix $\rho_{ij}^{\tau}$ may become singular if some single-particle states do not contribute to the total wave function, leading to numerical difficulties when evaluating the inverse $(\rho_{ij}^{\tau})^{-1}$. 
To solve this issue, we follow the method used to treat the overlap kernels in the gd-TDGCM \cite{LiB2023PRC}, and first diagonalize the one-body density matrix:
\begin{equation}
\rho^{\tau}_{ij}=\sum_kn_k^{\tau}\mathcal{U}_{ik}^{\tau}(\mathcal{U}_{kj}^{\tau})^{\dagger}.
\end{equation}
Only the terms with eigenvalues $n_k^{\tau}$ larger than a cutoff $\sigma$ are retained when computing the inverse of the one-body density matrix:
\begin{equation}
({\rho}^{\tau}_{ij})^{-1}\approx\sum_{n_k^{\tau}>\sigma}\left(\frac{1}{n_k^{\tau}}\right)\mathcal{U}_{ik}^{\tau}(\mathcal{U}_{kj}^{\tau})^{\dagger}.
\end{equation} 
In all the cases considered in this work, the eigenvalues of the one-body density matrix remained large throughout
the time evolution, so this truncation is not necessary.

The third modification concerns the numerical stability of the equation of motion for the expansion coefficients. 
Equation \eqref{EOM-C} is modified as
\begin{equation}
i\hbar\dot{C}_J=\sum_I\left(\mathcal{H}_{JI}-\mathcal{H}_{JI}^{\mathrm{MF}}-E\delta_{JI}\right)C_I.
\end{equation} 
Here, subtracting the total energy $E$ from the diagonal terms of the Hamiltonian kernel removes an overall oscillatory factor from the many-body wave function without affecting physical observables.
It improves the numerical stability of the coefficient evolution, particularly when many configurations with widely different energies are included.

\section{Numerical test}\label{Numerical-test}

\begin{figure*}[htbp!]
   \centering
   \includegraphics[width=0.6\linewidth]{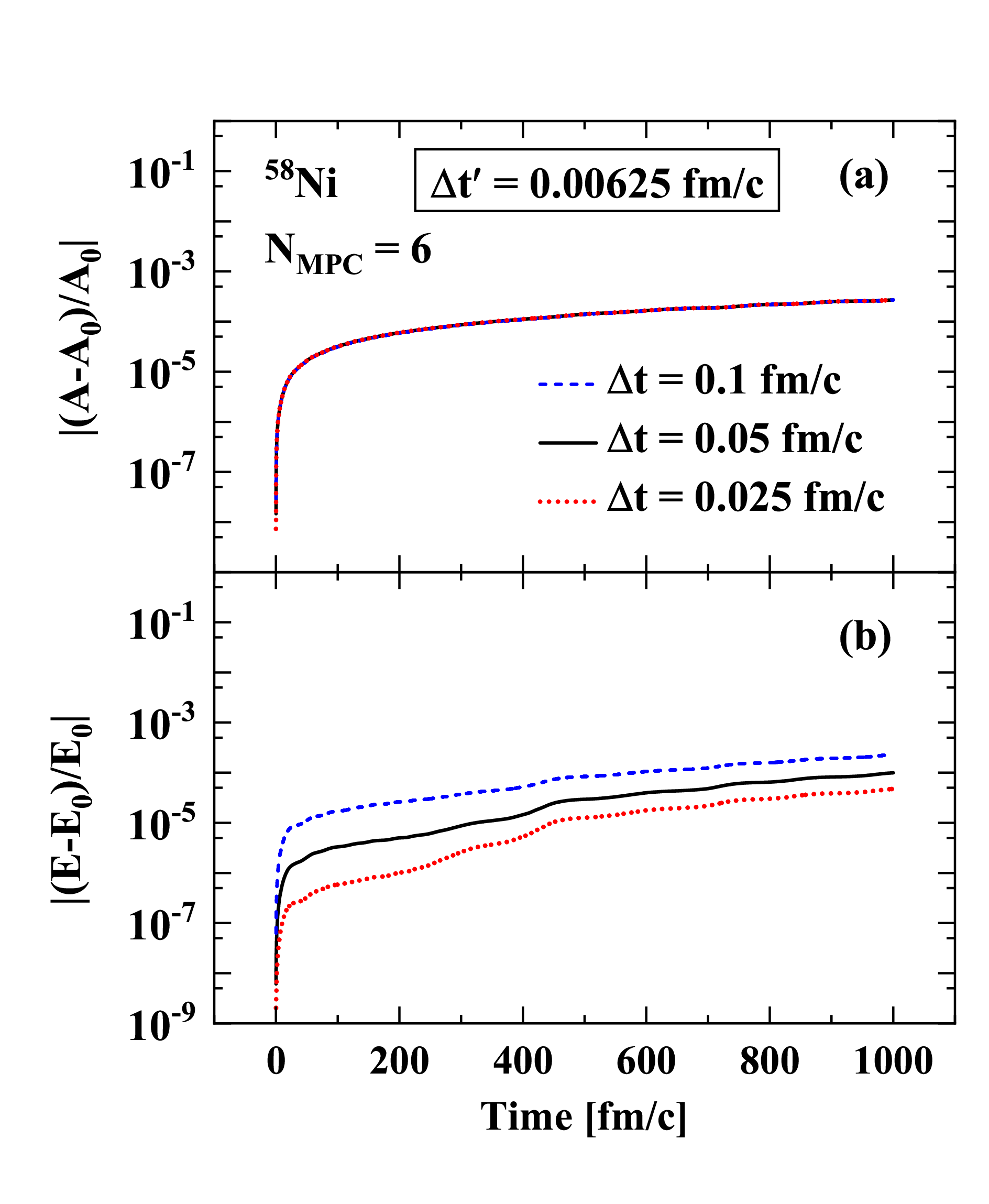}
   \caption{The relative particle number deviation $|(A-A_0)/A_0|$ (a), and the relative energy deviation $|(E-E_0)/E_0|$ (b), as functions of time, with respect to the initial particle number $A_0$ and initial energy $E_0$, respectively. The relative deviations are shown for $\Delta t = 0.1$, $0.05$, and $0.025$ fm/$c$, with fixed $\Delta t^{\prime}=0.00625$ fm/$c$.}
   \label{deltat-test}
 \end{figure*}

In this section, we perform a detailed numerical test of the conservation of energy and particle number, with respect to the choice of  the time steps $\Delta t$ and $\Delta t'$. The test case is the nucleus $^{58}$Ni with $N_{\text{MPC}}=6$ configurations. 
The explicit configuration space will be explained in detail in Sec. \ref{Results-and-discussions}.
Initially, the radius of $^{58}$Ni is constrained to 3.56 fm to start the monopole oscillation.
Figure \ref{deltat-test} displays the time evolution of the relative particle number deviation $|(A-A_0)/A_0|$, and the relative energy deviation $|(E-E_0)/E_0|$, for different values of $\Delta t = 0.1$, $0.05$, and $0.025$ fm/$c$, with fixed $\Delta t^{\prime}=0.00625$ fm/$c$.
In Fig. \ref{deltat-test} (a), one notices that the time evolution of the relative particle number deviation is stable and virtually independent of the choice of $\Delta t$.
The relative particle number deviation remains within $4\times 10^{-4}$ up to 1000 fm/$c$.
In contrast, the conservation of energy exhibits a consistent dependence on the value of $\Delta t$.
As shown in Fig. \ref{deltat-test} (b), the smaller $\Delta t$ is, the better the energy is conserved. 
These results can be understood as follows. 
The conservation of particle number is closely related to the orthonormality of single-particle states, whose evolution is governed primarily by $\Delta t'$ rather than $\Delta t$ (see Sec. \ref{Numerical-details}).
The total energy, on the other hand, is evaluated using the Hamiltonian kernels and expansion coefficients \eqref{observable-expectation}, and therefore, its accuracy is sensitive to $\Delta t$.

\begin{figure*}[htbp!]
   \centering
   \includegraphics[width=0.6\linewidth]{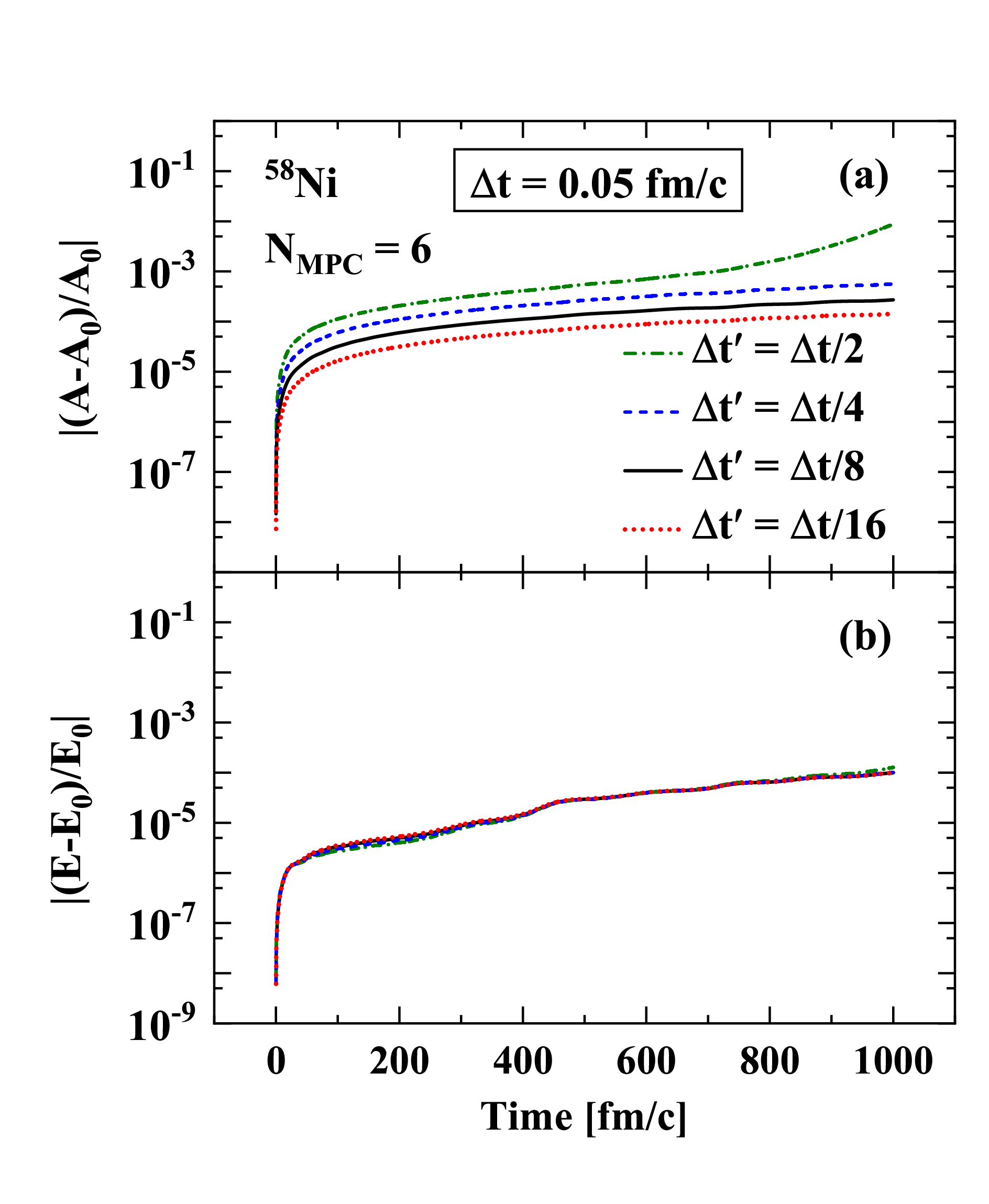}
   \caption{The relative particle number deviation $|(A-A_0)/A_0|$ (a), and the relative energy deviation $|(E-E_0)/E_0|$ (b), as functions of time, with respect to the initial particle number $A_0$ and initial energy $E_0$, respectively. The relative deviations are shown for $\Delta t^{\prime} = \Delta t/2$, $\Delta t/4$, $\Delta t/8$, and $\Delta t/16$, with fixed $\Delta t=0.05$ fm/$c$.}
   \label{deltatprime-test}
 \end{figure*}

To further illustrate the conservation of energy and particle number with respect to $\Delta t'$, Figure \ref{deltatprime-test} shows the time evolution of $|(A-A_0)/A_0|$ and $|(E-E_0)/E_0|$ for different $\Delta t'$ values, namely $\Delta t'= \Delta t/2$, $\Delta t/4$, $\Delta t/8$ and $\Delta t/16$, with fixed $\Delta t=0.05$ fm/$c$.
Not surprisingly, as shown in Fig. \ref{deltatprime-test}, the conservation of particle number improves as $\Delta t^{\prime}$ decreases, whereas the conservation of energy exhibits little dependence on $\Delta t^{\prime}$.
For $\Delta t = 0.05$ fm/$c$ and $\Delta t^{\prime}=\Delta t/8$, the relative energy deviation and particle number deviation are both within $4\times 10^{-4}$ up to 1000 fm/$c$.
This choice is therefore sufficiently accurate for the ISGMR calculations of $^{58}$Ni and $^{60}$Ni presented in the next section.
Similar conservation accuracy is obtained for the other $^{58}$Ni case with $N_{\text{MPC}}=15$ and for the $^{60}$Ni calculations, with the relative deviations of both quantities remaining within $4\times 10^{-4}$ over the same time interval.

\section{Results and discussion}
\label{Results-and-discussions}

\begin{figure*}[htbp!]
   \centering
   \includegraphics[width=0.95\linewidth]{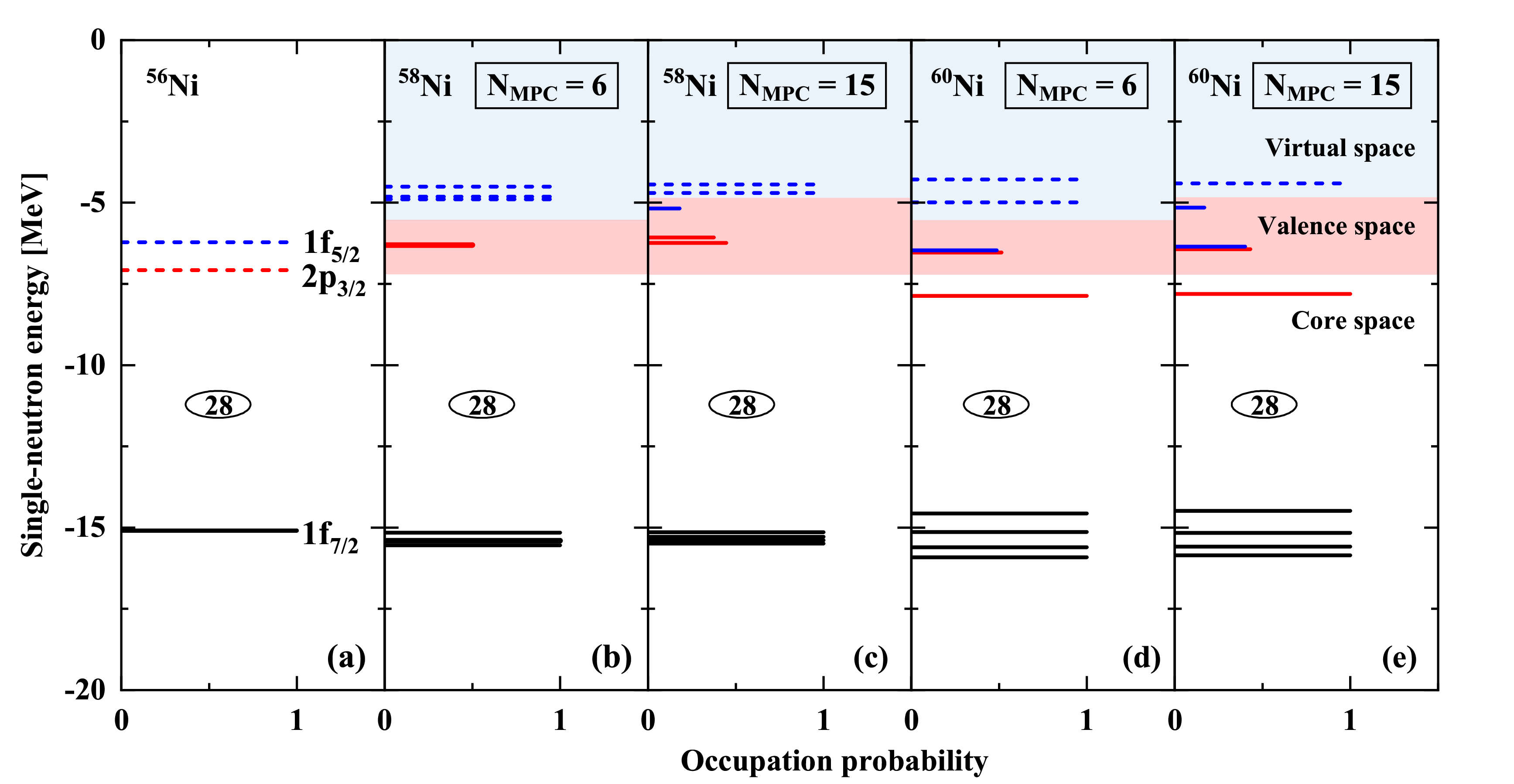}
   \caption{The RDFT-SLAP energies and occupation probabilities of single-neutron states at the initial time are shown for $^{58}$Ni in panels (b,c) and for $^{60}$Ni in (d,e). For reference, the single-neutron states of the ground state of the spherical nucleus $^{56}$Ni are shown on the left in panel (a), with energies shifted up by 2.0 MeV.}
   \label{init-state}
 \end{figure*}

To start monopole density oscillations in $^{58}$Ni and $^{60}$Ni, the initial wave functions, i.e., both the expansion coefficients and the single-particle states, are obtained from RDFT-SLAP calculations with a constrained radius (see Appendix D). 
Initially, we compress the equilibrium radius by about 0.1 fm. 
Here, the constrained radius is 3.56 fm for $^{58}$Ni and 3.64 fm for $^{60}$Ni, respectively. 
The resulting energies and occupation probabilities of the initial single-neutron states are shown in Fig. \ref{init-state}. 
The occupation probability $n_{i,\tau}$ of the single-particle state $\phi_{i,\tau}$ is evaluated from the RDFT-SLAP expansion coefficients as,
\begin{equation}\label{n-i-tau}
  n_{i,\tau}=\sum_I|C_I|^2P_I^{i,\tau}, 
  ~~P_I^{i,\tau}=\left\{ \begin{aligned}
                         &1, \text{~~} \phi_{i,\tau} \text{ is occupied in~} |\text{MPC}_I\rangle; \\
                        &0, \text{~~otherwise}. 
                      \end{aligned}\right.
\end{equation}
Because time-reversal symmetry is preserved in the initial RDFT-SLAP solutions, the single-neutron states are two-fold degenerate.
For $^{58}$Ni, the valence space consists of the four $2p_{3/2}$ states above the $N=28$ closed shell for $N_{\text{MPC}}=6$, and additionally the two lowest $1f_{5/2}$ states for $N_{\text{MPC}}=15$.
For $^{60}$Ni, the valence space consists of two $2p_{3/2}$ and two $1f_{5/2}$ states for $N_{\text{MPC}}=6$, and additionally two more $1f_{5/2}$ states for $N_{\text{MPC}}=15$.
A larger calculation with $N_{\text{MPC}}=70$, discussed below, is obtained by distributing the four valence neutrons of $^{60}$Ni over the eight lowest single-particle states above the $N=28$ closed shell, yielding $C_8^4=70$ configurations.
The $N_{\text{MPC}}=70$ space provides a useful check of the stability of the monopole response with respect to an enlarged valence space.
 
It should be emphasized that the core states are not kept frozen during the time evolution. 
They evolve dynamically according to Eq. \eqref{phi-a}, which reduces to the standard TDDFT equation \cite{Ren2020PLB,RenZX2020PRC} when the valence space is empty and pairing correlations are neglected.
Since TDDFT is equivalent to RPA in the small-amplitude limit \cite{Ring2004The}, the evolution of the core states corresponds to the usual RPA response.
Configuration mixing in the valence space then introduces correlations beyond the conventional RPA description of collective motion.

\begin{figure*}[htbp!]
   \centering
   \includegraphics[width=0.95\linewidth]{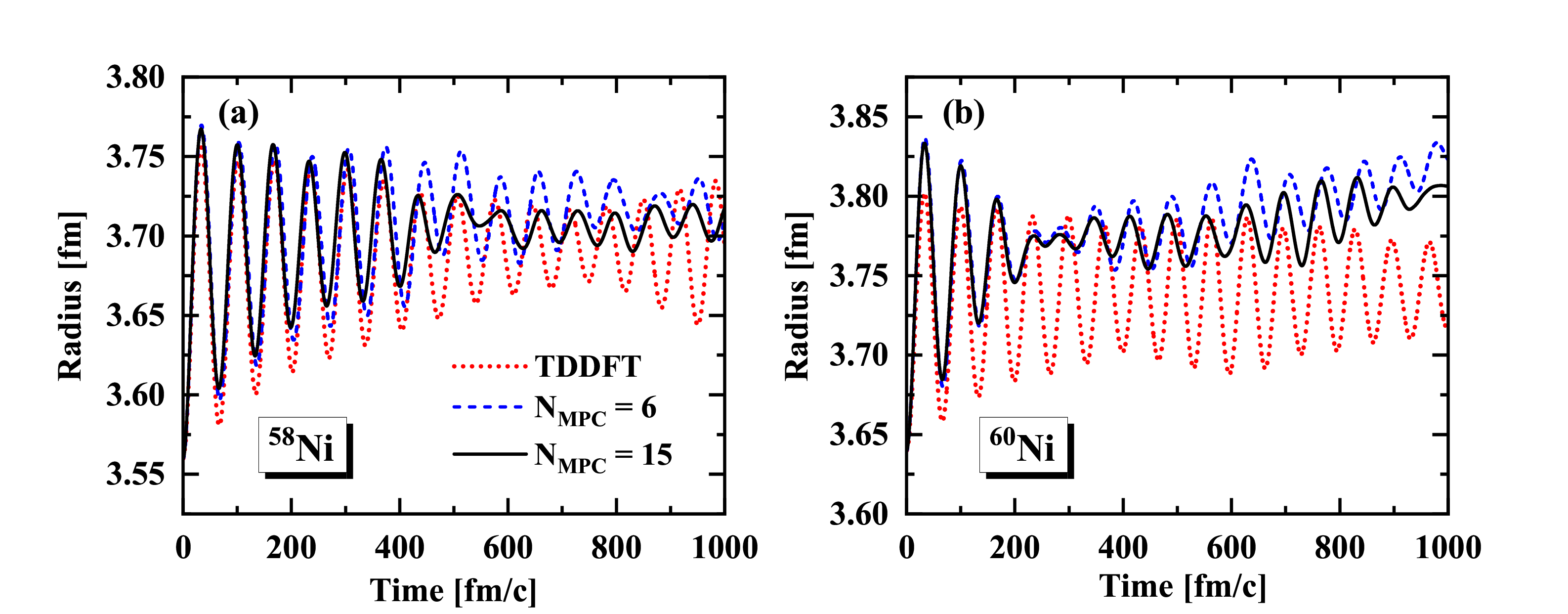}
   \caption{Time evolution of the radius for $^{58}$Ni (a) and $^{60}$Ni (b) calculated with CI-TDDFT using $N_{\text{MPC}}=6$ (blue dashed curve) and $N_{\text{MPC}}=15$ (black solid curve) valence configurations, in comparison with the TDDFT results (red dotted curve).}
   \label{radius-Ni58-Ni60}
 \end{figure*}

Figure \ref{radius-Ni58-Ni60} depicts the time evolution of the radius for $^{58}$Ni and $^{60}$Ni calculated with CI-TDDFT using $N_{\text{MPC}}=6$ and $N_{\text{MPC}}=15$ valence configurations.
For comparison, we also present the TDDFT results, with pairing correlations treated dynamically within the Bardeen-Cooper-Schrieffer (BCS) approximation \cite{RenZX2022PRC2}.
The TDDFT calculations are performed on the same 3D lattice space as in CI-TDDFT, using the same PC-PK1 functional in the particle-hole channel and a monopole pairing interaction in the particle-particle channel.
Proton pairing vanishes, while the neutron pairing strengths are 0.7 MeV for $^{58}$Ni and 0.63 MeV for $^{60}$Ni, which are adjusted by reproducing the experimental odd-even mass differences. 
For $^{58}$Ni, the CI-TDDFT and TDDFT radii exhibit similar oscillatory patterns at early times.
At later times, especially after about $400$ fm/$c$, the CI-TDDFT trajectories show a more complex oscillatory pattern than the TDDFT result.
This behavior indicates that additional frequency components are generated by the configuration mixing as time evolves.
For $^{60}$Ni, the difference between CI-TDDFT and TDDFT appears already at the beginning of the time evolution, exhibiting a stronger sensitivity of the monopole dynamics to the initial configuration mixing in the valence space.

From the time evolution of the radius, the ISGMR strength distributions are extracted following the procedure described in Appendix E and shown in Fig. \ref{strength-Ni58-Ni60}.
The TDDFT results are already in good agreement with the experimental excitation energies of the ISGMR in $^{58}$Ni ($18.43\pm 0.15$ MeV) and $^{60}$Ni ($17.62\pm 0.15$ MeV) \cite{LiuYW2006PRC}.
However, because TDDFT reduces to the RPA in the small-amplitude limit, it does not include higher-order configuration mixing that contributes to the spreading width of giant resonances.
In contrast, CI-TDDFT produces visibly broader strength distributions, while the main peak positions remain close to the TDDFT values within the resolution of the present calculation.
For both nuclei, enlarging the configuration space from $N_{\text{MPC}}=6$ to $N_{\text{MPC}}=15$ further broadens the calculated strength distributions.
The effect of configuration mixing when compared with TDDFT is more pronounced for $^{60}$Ni, for which the CI-TDDFT strength
distribution displays a two-peak structure. 
This structure is consistent with the more complex time evolution of the radius shown in Fig. \ref{radius-Ni58-Ni60} (b).
A direct comparison with the experimental total widths, $7.41 \pm 0.13$ MeV for $^{58}$Ni and $7.55 \pm 0.13$ MeV for $^{60}$Ni \cite{LiuYW2006PRC}, should, however, be regarded as qualitative at this stage.
The present CI-TDDFT calculation mainly accounts for the spreading width generated by configuration mixing in the
adopted valence space, whereas other contributions to the experimental width, such as escape effects and other damping mechanisms, are not included.

\begin{figure*}[htbp!]
   \centering
   \includegraphics[width=0.45\linewidth]{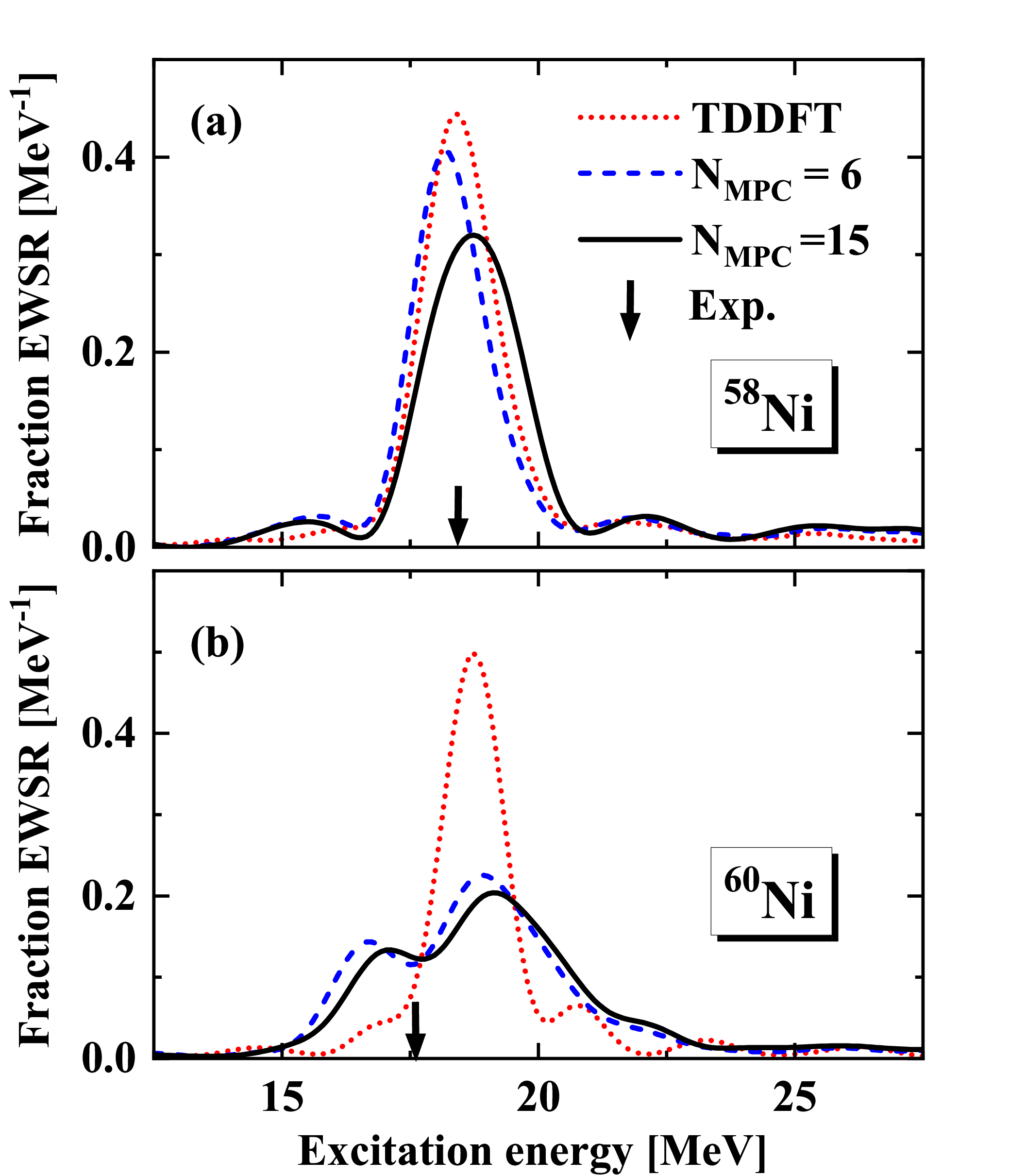}
   \caption{Fraction of the ISGMR energy-weighted sum rule (EWSR) as a function of the excitation energy for $^{58}$Ni (a) and $^{60}$Ni (b) computed with CI-TDDFT using $N_{\text{MPC}}=6$ (blue dashed curve) and $N_{\text{MPC}}=15$ (black solid curve) valence configurations, in comparison with the TDDFT results (red dotted curve).}
   \label{strength-Ni58-Ni60}
 \end{figure*}

\begin{figure*}[htbp!]
   \centering
   \includegraphics[width=0.8\linewidth]{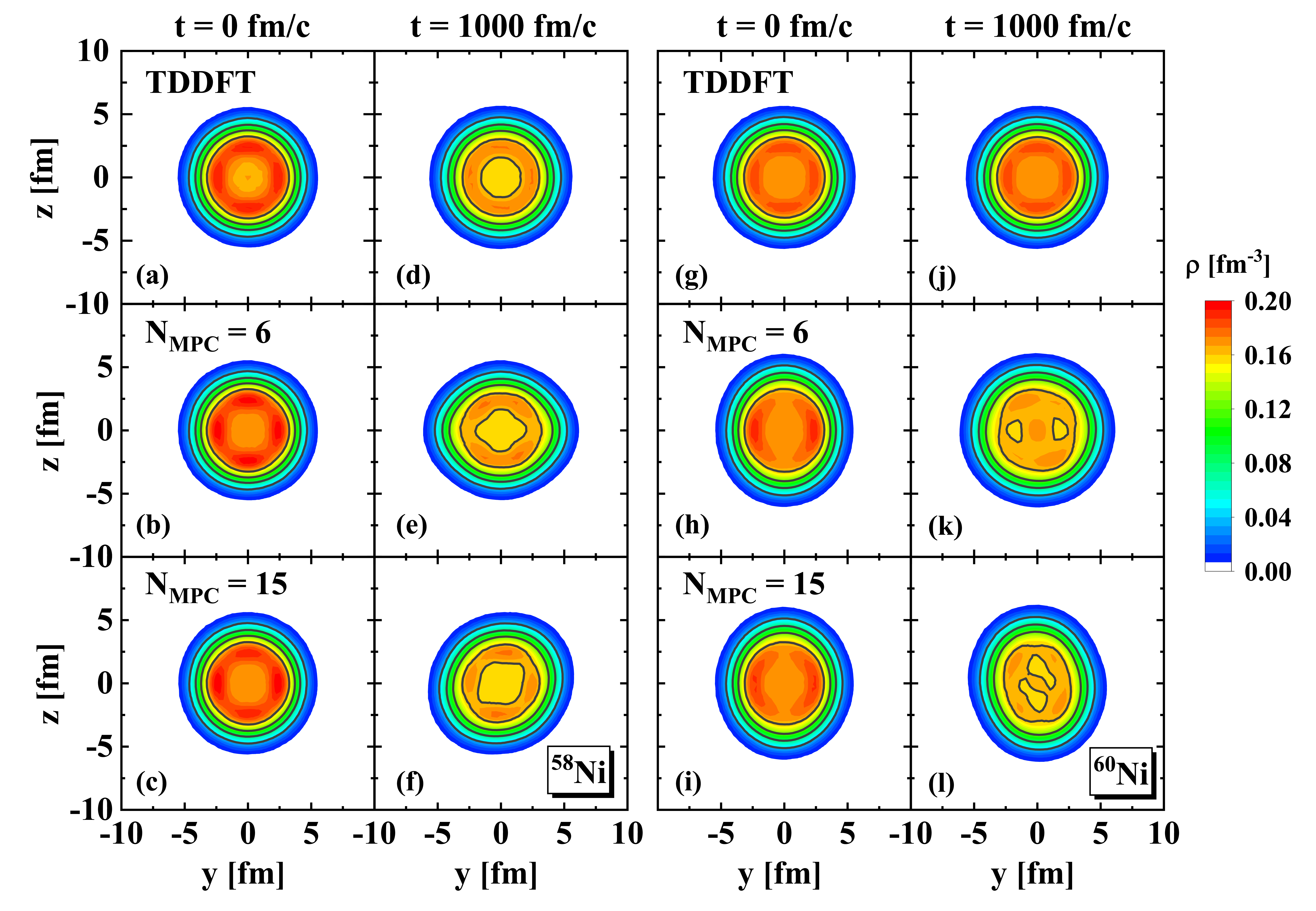}
   \caption{Density profile of $^{58}$Ni (a-f) and $^{60}$Ni (g-l) in the $x=0$ plane at $t =$ 0.0 fm/$c$ and 1000 fm/$c$ calculated with CI-TDDFT using $N_{\text{MPC}}=6$ (middle) and $N_{\text{MPC}}=15$ (bottom) valence configurations, in comparison with the TDDFT results (top).} 
   \label{density-profile}
 \end{figure*}

The density distributions in 3D coordinate space provide further insight into the difference between the single Slater determinant used in TDDFT and the correlated wave function of the CI-TDDFT.
Figure \ref{density-profile} shows the density profiles of $^{58}$Ni and $^{60}$Ni in the $x=0$ plane at $t =$ 0.0 fm/$c$ and 1000 fm/$c$, calculated with CI-TDDFT using $N_{\text{MPC}}=6$ and $N_{\text{MPC}}=15$ valence configurations, in comparison with TDDFT results. The density is calculated as
\begin{equation}
\rho(\boldsymbol{r},t)=\langle\Psi(t)|d_{\boldsymbol{r}}^{\dagger}d_{\boldsymbol{r}}|\Psi(t)\rangle,
\end{equation}  
where $d_{\boldsymbol{r}}^{\dagger}$ creates a nucleon at the lattice point ${\boldsymbol{r}}$, and $\Psi(t)$ denotes either the  correlated CI-TDDFT wave function or the TDDFT Slater determinant.
In the TDDFT calculations, the density distribution of $^{58}$Ni remains spherical during the time evolution because the initial spherical symmetry is preserved self-consistently.
The initial CI-TDDFT densities for $^{58}$Ni are also nearly spherical.
During the subsequent time evolution, however, configuration mixing induces non-spherical components in the density distribution, including quadrupole and higher-order deformations.
This suggests that the broadened CI-TDDFT strength distribution is associated with the coupling of the monopole oscillation to additional collective degrees of freedom.
This interpretation is analogous to the mechanism discussed in the gd-TDGCM \cite{LiB2023PRC}, where quantum fluctuations in the collective coordinate space can broaden the response.
For $^{60}$Ni, the CI-TDDFT densities show a pronounced non-spherical component already at the initial time, reflecting the stronger initial configuration mixing in this nucleus.

\begin{figure*}[htbp!]
   \centering
   \includegraphics[width=0.8\linewidth]{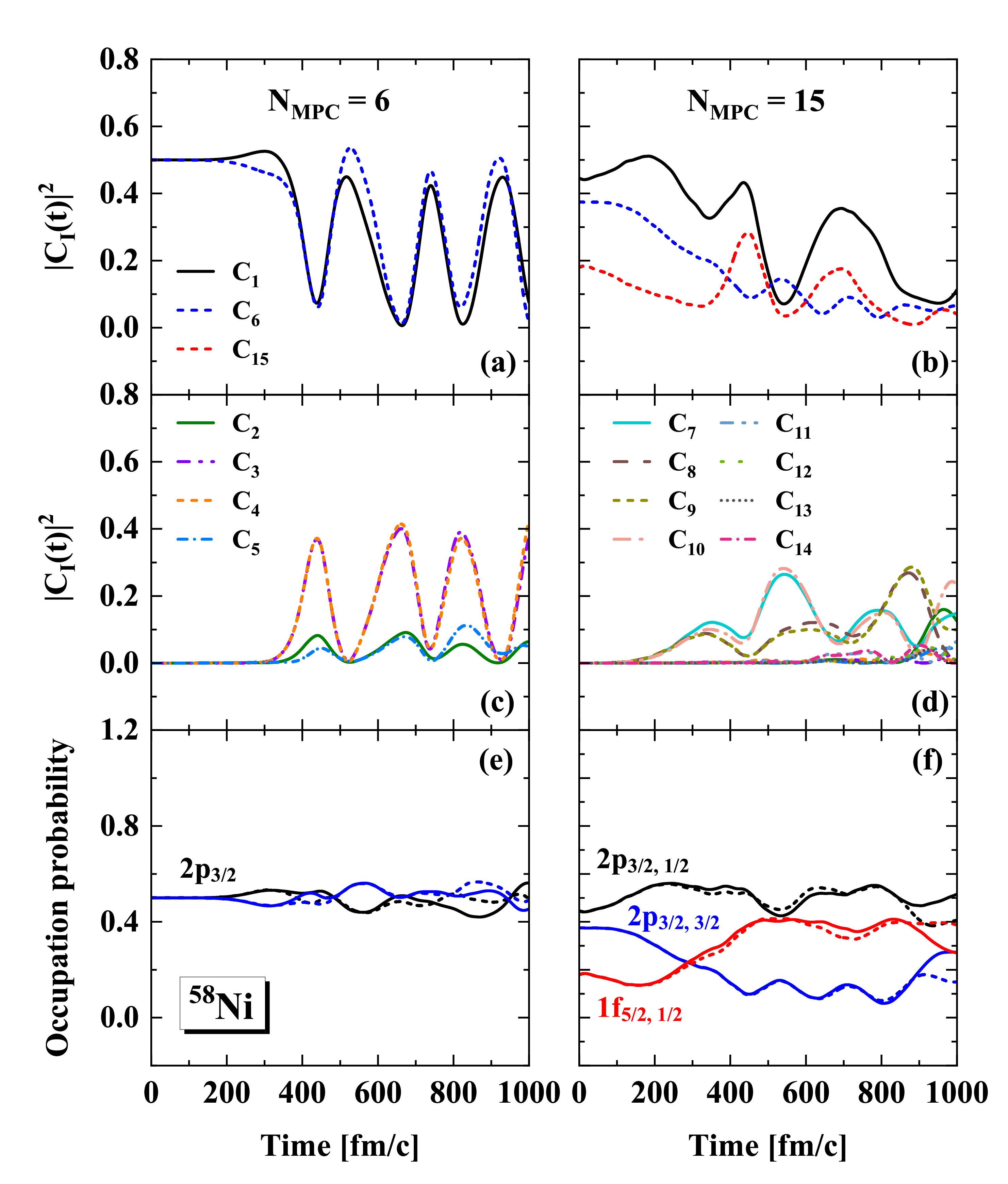}
   \caption{Time evolution of the square moduli of the expansion coefficients $|C_I(t)|^2$ (top and middle) and the occupation probabilities of the valence single-neutron states (bottom) for $^{58}$Ni computed with CI-TDDFT using $N_{\text{MPC}}=6$ (left) and $N_{\text{MPC}}=15$ (right) valence configurations. The configurations are sorted by their energies from lowest to highest, and are denoted as $|\text{MPC}_I\rangle$, where $I=1, 2, \cdots$. The valence single-neutron states are labeled by the spherical quantum numbers of their main component at the initial time.}
   \label{Coef-occ-Ni58}
 \end{figure*}

To analyze the role of individual configurations, we calculate the squared moduli of the expansion coefficients, $|C_I(t)|^2$, which quantify the contribution of the configuration $|\text{MPC}_I\rangle$ to the total wave function at time $t$.
As an example, Fig. \ref{Coef-occ-Ni58} shows the time evolution of $|C_I(t)|^2$ and the occupation probabilities of the valence single-neutron states for $^{58}$Ni.
The configurations are sorted by their energies from lowest to highest and are denoted by $|\text{MPC}_I\rangle$ with $I=1, 2, \cdots$.
For the $N_{\text{MPC}}=6$ case, the initial wave function contains only the two fully-paired configurations $I=1$ and $I=6$, which carry equal weight.
After about 300 fm/$c$, their weights start to oscillate in phase with a period of 200 fm/$c$.  
At the same time, the configurations $I=2,3,4,5$, which have zero initial amplitudes, acquire appreciable probabilities.
Their out-of-phase behavior with respect to $I=1$ and $I=6$ reflects the redistribution of probability among the available configurations during the collective motion.
Overall, the occupation probabilities of the four $2p_{3/2}$ states remain around $50\%$ throughout the time evolution.
For the $N_{\text{MPC}}=15$ case, the configuration weights evolve in a more complex pattern.
The initial wave function contains three fully-paired configurations, $I=1$, $6$, and $15$, and their weights are displayed in Fig. \ref{Coef-occ-Ni58} (b).
The initially zero-weight configurations $I=7$-$10$ acquire non-negligible contributions already at early times.
The corresponding single-particle occupation probabilities show that the $2p_{3/2,3/2}$ and $2p_{3/2,1/2}$ states dominate initially.
After about $300$ fm/$c$, the two $1f_{5/2,1/2}$ states become more important than the $2p_{3/2,3/2}$ states.
In contrast, in a TDDFT description of $^{58}$Ni based on a single Slater determinant, since these $1f_{5/2,1/2}$ levels are always unoccupied, these $1f_{5/2,1/2}$ components would be absent in the total wavefunction during the time evolution. 
Therefore, this provides a direct illustration of the additional valence-space correlations included in CI-TDDFT.

\begin{figure*}[htbp!]
   \centering
   \includegraphics[width=0.45\linewidth]{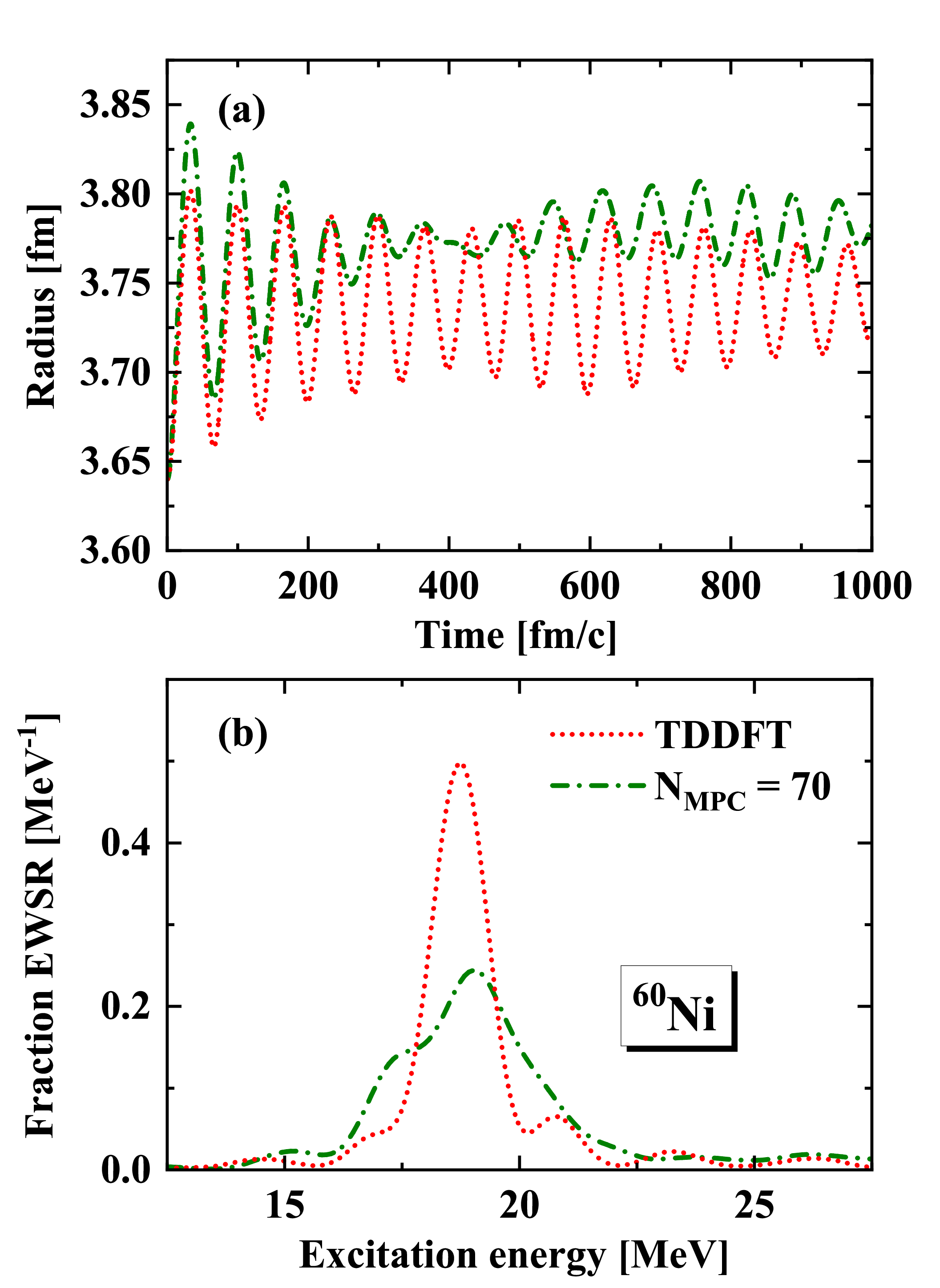}
   \caption{Time evolution of the radius (a) and ISGMR strength distribution (b) for $^{60}$Ni calculated with CI-TDDFT using $N_{\text{MPC}}=70$ (dash-dotted curve) valence configurations, in comparison with the TDDFT results (red dotted curve).}
   \label{radius-strength-NMPC70}
 \end{figure*}

Finally, we have performed a larger calculation for $^{60}$Ni with $N_{\text{MPC}} = 70$. 
In this case, the four valence neutrons are distributed over the eight lowest single-particle states above the $N=28$ closed shell. The neutron pairing strength is adjusted to $G^{\text{n}}=1.2$ MeV in this enlarged space, in order to reproduce the experimental
odd-even mass difference. 
To make the calculation computationally feasible, larger time steps are used, namely $\Delta t=0.1$ fm/$c$ and $\Delta t'=\Delta t/4$. 
With this choice, the relative energy deviation and particle-number deviation remain of the order of $10^{-3}$ up to $1000$ fm/$c$.
As shown in Fig.~\ref{radius-strength-NMPC70}, the ISGMR strength distribution obtained with $N_{\rm MPC}=70$ is similar to those calculated in the smaller spaces with $N_{\rm MPC}=6$ and $N_{\rm MPC}=15$. 
This indicates that, for the monopole response considered here, the additional low-lying $2p_{3/2}$ states included in the enlarged valence space have only a minor effect on the calculated strength distribution, and therefore, the present ISGMR observable is relatively stable with respect to this particular enlargement of the valence space.

\section{Summary and outlook}\label{summary-and-outlook}

We have developed and implemented a configuration-interaction time-dependent density functional theory (CI-TDDFT) for nuclear dynamics. The central idea of the method is to expand the correlated nuclear many-body wave function in terms of time-dependent many-particle configurations constructed from a common set of orthonormal single-particle basis states. In contrast to conventional TDDFT, which is based on a single product state, the present framework incorporates configuration mixing into the many-body wave function, thereby providing a fully quantum approach to include beyond-mean-field correlations.

Using the Dirac–Frenkel time-dependent variational principle, we have derived coupled equations of motion for both the expansion coefficients and the single-particle states. Because the variational procedure is applied consistently to both sets of degrees of freedom, the resulting model for nuclear dynamics conserves the total energy of the system. Moreover, since each many-particle configuration has a fixed number of neutrons and protons, particle number is conserved even in the presence of pairing correlations.

The nuclear Hamiltonian in the current implementation includes the relativistic density functional PC-PK1 in the particle-hole channel and a monopole pairing interaction in the particle-particle channel. For practical implementation, the single-particle Hilbert space is divided into core, valence, and virtual spaces, with configuration mixing restricted to the valence space.

The numerical implementation has been tested by modeling the time evolution of $^{58}$Ni. 
The impact of the time-step choice for propagating the expansion coefficients and the single-particle states was examined in detail.
Using a time step of $\Delta t=0.05$ fm/$c$ for the expansion coefficients and $\Delta t'=\Delta t/8$ for the single-particle states,
the relative deviations in both total energy and particle number remain within $4\times 10^{-4}$ up to a propagation time of 1000 fm/$c$. These results demonstrate the numerical stability of our implementation and confirm the conservation of energy and particle number expected from the variational formulation.

As a first illustrative application, we have calculated the isoscalar giant monopole resonances in  $^{58}$Ni and $^{60}$Ni. The initial states are generated from stationary self-consistent RDFT-SLAP calculations by imposing constraints on the nuclear radii. The subsequent time evolution is modeled using CI-TDDFT and compared with corresponding TDDFT results. For the giant monopole resonance, CI-TDDFT generally yields broader strength distributions while keeping the main peak positions close to those obtained from TDDFT. This broadening becomes more pronounced as the valence configuration space is enlarged from $N_{\rm MPC}=6$ to $N_{\rm MPC}=15$.

An analysis of density distributions and configuration amplitudes provides a microscopic interpretation of this effect. In TDDFT, the dynamics is restricted to the evolution of a single Slater determinant, and the densities of  $^{58}$Ni and $^{60}$Ni remain spherical during the monopole oscillation. In contrast, CI-TDDFT introduces configuration mixing, which induces non-spherical components in the density and leads to a redistribution of probabilities among different many-particle configurations. This indicates that the monopole mode couples to additional collective degrees of freedom via valence-space correlations. The effect is particularly pronounced in $^{60}$Ni, where the calculated CI-TDDFT strength distribution exhibits a two-peak structure. A further calculation for $^{60}$Ni with $N_{\rm MPC}=70$ configurations shows that the monopole response does not differ significantly from the results obtained with smaller configuration spaces.

The model developed in this study represents an important step toward a fully microscopic, quantum configuration-interaction extension of nuclear TDDFT. The comparison with experimental total widths should still be regarded as qualitative, as the current model accounts only for the spreading width arising from configuration mixing within the adopted valence space, while the escape width and other damping effects are not included. Future work will focus on larger configuration spaces, more optimized way to obtain the initial state, and applications to other collective modes as well as large-amplitude nuclear dynamical processes. In particular, the CI-TDDFT framework provides a promising basis for investigating the role of many-body correlations, quantum fluctuations, and configuration mixing in nuclear fission, fusion, and multinucleon transfer reactions.

\appendix

\section{One-body and two-body density matrices}\label{appendixA}

The one-body density matrix reads
\begin{equation}
\begin{split}
\rho_{ij}^{\tau} &= \langle\Psi| c_{i,\tau}^{\dagger} c_{j,\tau} |\Psi\rangle \\
&= \sum_{J,I} C_J^* C_I \langle \text{MPC}_J | c_{i,\tau}^{\dagger} c_{j,\tau} | \text{MPC}_I \rangle,
\end{split}
\end{equation}
where the matrix elements $\langle \text{MPC}_J | c_{i,\tau}^{\dagger} c_{j,\tau} | \text{MPC}_I \rangle$ are evaluated as follows. Considering the occupied single-particle states in $|\text{MPC}_I\rangle$ and $|\text{MPC}_J\rangle$, two cases arise:

(a) $|\text{MPC}_I\rangle$ and $|\text{MPC}_J\rangle$ are identical: $|\text{MPC}_J\rangle = |\text{MPC}_I\rangle$. Then,
\begin{equation}
\langle \cdots | c_{i,\tau}^{\dagger} c_{j,\tau} | \cdots \rangle = \delta_{ij}I_{i,\tau}.
\end{equation}
Here, $\{I_{i,\tau}=0,1\}$ are the occupation numbers of the single-particle state $|\phi_{i,\tau}\rangle$ in $|\text{MPC}_I\rangle$.

(b) $|\text{MPC}_I\rangle$ differs from $|\text{MPC}_J\rangle$ by the occupation of a single state. Specifically, suppose a single-particle state $|\phi_{\lambda,\tau_0}\rangle$ is occupied in $|\text{MPC}_J\rangle$ but unoccupied in $|\text{MPC}_I\rangle$, and a state $|\phi_{\alpha,\tau_0}\rangle$ is unoccupied in $|\text{MPC}_J\rangle$ but occupied in $|\text{MPC}_I\rangle$. Then
\begin{equation}
\langle \cdots 1_{\lambda,\tau_0}\cdots 0_{\alpha,\tau_0}\cdots | c_{i,\tau}^{\dagger} c_{j,\tau} | \cdots 1_{\alpha,\tau_0}\cdots 0_{\lambda,\tau_0}\cdots \rangle = (-1)^{\mathcal{P}} \delta_{j\alpha}\delta_{i\lambda}\delta_{\tau\tau_0},
\end{equation}
where the phase factor $\mathcal{P}$ is
\begin{equation}
\mathcal{P} = \sum_{k=1}^{\lambda-1} I_{k,\tau_0} + \sum_{k=1}^{\alpha-1} I'_{k,\tau_0}.
\end{equation}
Here, $\{I_{k,\tau}=0,1\}$ are the occupation numbers of single-particle states in $|\text{MPC}_J\rangle$, and $\{I'_{k,\tau}=0,1\}$ those in $|\text{MPC}_I\rangle$.

The two-body density matrix for the particle-hole channel reads 
\begin{equation}
\begin{split}
\rho_{ijkl}^{\tau\tau^{\prime}}=&\langle\Psi|(c_{i,\tau}^{\dagger}c_{j,\tau})(c_{k,\tau^{\prime}}^{\dagger}c_{l,\tau^{\prime}})|\Psi\rangle\\
=&\sum_{J,I}C_J^*C_I\langle\text{MPC}_J|(c_{i,\tau}^{\dagger}c_{j,\tau})(c_{k,\tau^{\prime}}^{\dagger}c_{l,\tau^{\prime}})|\text{MPC}_I\rangle.
\end{split}
\end{equation}
The matrix elements $\langle\text{MPC}_J|(c_{i,\tau}^{\dagger}c_{j,\tau})(c_{k,\tau^{\prime}}^{\dagger}c_{l,\tau^{\prime}})|\text{MPC}_I\rangle$ are evaluated as follows. 
It should be emphasized that the Hartree approximation is adopted for the particle-hole channel, which means that Wick contractions between operators from different brackets are neglected.
Four cases arise:

(a) $|\text{MPC}_I\rangle$ and $|\text{MPC}_J\rangle$ are identical: 
\begin{equation}
\langle\cdots|(c_{i,\tau}^{\dagger}c_{j,\tau})(c_{k,\tau^{\prime}}^{\dagger}c_{l,\tau^{\prime}})|\cdots\rangle=\delta_{ij}\delta_{kl}I_{i,\tau}I_{k,\tau^{\prime}}.
\end{equation}

(b) $|\text{MPC}_I\rangle$ differs from $|\text{MPC}_J\rangle$ by the occupation of a single state:
\begin{equation}
\begin{split}
&\langle\cdots 1_{\lambda,\tau_0}\cdots 0_{\alpha,\tau_0}\cdots|(c_{i,\tau}^{\dagger}c_{j,\tau})(c_{k,\tau^{\prime}}^{\dagger}c_{l,\tau^{\prime}})|\cdots 1_{\alpha,\tau_0}\cdots 0_{\lambda,\tau_0}\cdots\rangle\\
=&(-1)^{\mathcal{P}}(\delta_{ij}\delta_{k\lambda}\delta_{l\alpha}\delta_{\tau_0\tau^{\prime}}+\delta_{kl}\delta_{i\lambda}\delta_{j\alpha}\delta_{\tau_0\tau}),
\end{split}
\end{equation}
where $\mathcal{P}=\sum_{k=1}^{\lambda-1}I_{k,\tau_0}+\sum_{k=1}^{\alpha-1}I_{k,\tau_0}^{\prime}$. 

(c) $|\text{MPC}_I\rangle$ differs from $|\text{MPC}_J\rangle$ by the occupation of two neutron (proton) states. Specifically, suppose two single-neutron (proton) states $|\phi_{\zeta,\tau_0}\rangle$ and $|\phi_{\xi,\tau_0}\rangle$ are occupied in $|\text{MPC}_J\rangle$ but unoccupied in $|\text{MPC}_I\rangle$, while two states $|\phi_{\alpha,\tau_0}\rangle$ and $|\phi_{\lambda,\tau_0}\rangle$ are unoccupied in $|\text{MPC}_J\rangle$ but occupied in $|\text{MPC}_I\rangle$. Then, 
\begin{equation}
\begin{split}
&\langle\cdots 1_{\zeta,\tau_0}\cdots 1_{\xi,\tau_0}\cdots 0_{\lambda,\tau_0}\cdots 0_{\alpha,\tau_0}\cdots|(c_{i,\tau}^{\dagger}c_{j,\tau})(c_{k,\tau^{\prime}}^{\dagger}c_{l,\tau^{\prime}})|\cdots 1_{\alpha,\tau_0}\cdots 1_{\lambda,\tau_0}\cdots 0_{\xi,\tau_0}\cdots 0_{\zeta,\tau_0}\cdots\rangle\\
=&(-1)^{\mathcal{P}}(-\delta_{i\xi}\delta_{k\zeta}\delta_{j\lambda}\delta_{l\alpha}+\delta_{i\xi}\delta_{k\zeta}\delta_{j\alpha}\delta_{l\lambda}+\delta_{i\zeta}\delta_{k\xi}\delta_{j\lambda}\delta_{l\alpha}-\delta_{i\zeta}\delta_{k\xi}\delta_{j\alpha}\delta_{l\lambda})\delta_{\tau\tau_0}\delta_{\tau^{\prime}\tau_0},
\end{split}
\end{equation}
where $\mathcal{P}=\sum_{k=1}^{\zeta-1}I_{k,\tau_0}+\sum_{k=1}^{\xi-1}I_{k,\tau_0}+\sum_{k=1}^{\alpha-1}I_{k,\tau_0}^{\prime}+\sum_{k=1}^{\lambda-1}I_{k,\tau_0}^{\prime}$.

(d) $|\text{MPC}_I\rangle$ differs from $|\text{MPC}_J\rangle$ by the occupation of one neutron and one proton state: 
\begin{equation}
\begin{split}
&\langle\cdots 1_{\zeta,\tau_2}\cdots 0_{\xi,\tau_2}\cdots 1_{\lambda,\tau_1}\cdots 0_{\alpha,\tau_1}\cdots|(c_{i,\tau}^{\dagger}c_{j,\tau})(c_{k,\tau^{\prime}}^{\dagger}c_{l,\tau^{\prime}})|\cdots 1_{\alpha,\tau_1}\cdots 0_{\lambda,\tau_1}\cdots 1_{\xi,\tau_2}\cdots 0_{\zeta,\tau_2}\cdots\rangle\\
=&(-1)^{\mathcal{P}}(\delta_{i\lambda}\delta_{j\alpha}\delta_{k\zeta}\delta_{l\xi}\delta_{\tau\tau_1}\delta_{\tau^{\prime}\tau_2}+\delta_{i\zeta}\delta_{j\xi}\delta_{k\lambda}\delta_{l\alpha}\delta_{\tau\tau_2}\delta_{\tau^{\prime}\tau_1}),
\end{split}
\end{equation}
where $\mathcal{P}=\sum_{k=1}^{\zeta-1}I_{k,\tau_2}+\sum_{k=1}^{\lambda-1}I_{k,\tau_1}+\sum_{k=1}^{\alpha-1}I_{k,\tau_1}^{\prime}+\sum_{k=1}^{\xi-1}I_{k,\tau_2}^{\prime}$, and $\tau_1\neq \tau_2$.

The two-body density matrix for the particle-particle channel reads 
\begin{equation}
\begin{split}
\kappa_{ijkl}^{\tau}&=\langle\Psi|c_{i,\tau}^{\dagger}c_{j,\tau}^{\dagger}c_{k,\tau}c_{l,\tau}|\Psi\rangle\\
&=\sum_{J,I}C_J^*C_I\langle\text{MPC}_J|c_{i,\tau}^{\dagger}c_{j,\tau}^{\dagger}c_{k,\tau}c_{l,\tau}|\text{MPC}_I\rangle.
\end{split}
\end{equation}
The matrix elements $\langle\text{MPC}_J|c_{i,\tau}^{\dagger}c_{j,\tau}^{\dagger}c_{k,\tau}c_{l,\tau}|\text{MPC}_I\rangle$ are evaluated for the three possible cases:

(a) $|\text{MPC}_I\rangle$ and $|\text{MPC}_J\rangle$ are identical: 
\begin{equation}
\langle\cdots|c_{i,\tau}^{\dagger}c_{j,\tau}^{\dagger}c_{k,\tau}c_{l,\tau}|\cdots\rangle=(\delta_{jk}\delta_{il}-\delta_{ik}\delta_{jl})I_{i,\tau}I_{j,\tau}.
\end{equation}

(b) $|\text{MPC}_I\rangle$ differs from $|\text{MPC}_J\rangle$ by the occupation of a single state:
\begin{equation}
\begin{split}
&\langle\cdots 1_{\lambda,\tau_0}\cdots 0_{\alpha,\tau_0}\cdots|c_{i,\tau}^{\dagger}c_{j,\tau}^{\dagger}c_{k,\tau}c_{l,\tau}|\cdots 1_{\alpha,\tau_0}\cdots 0_{\lambda,\tau_0}\cdots\rangle\\
=&(-1)^{\mathcal{P}}(\delta_{jk}\delta_{i\lambda}\delta_{l\alpha}+\delta_{il}\delta_{j\lambda}\delta_{k\alpha}-\delta_{ik}\delta_{j\lambda}\delta_{l\alpha}-\delta_{jl}\delta_{i\lambda}\delta_{k\alpha})\delta_{\tau\tau_0},
\end{split}
\end{equation}
where $\mathcal{P}=\sum_{k=1}^{\lambda-1}I_{k,\tau_0}+\sum_{k=1}^{\alpha-1}I_{k,\tau_0}^{\prime}$.

(c) $|\text{MPC}_I\rangle$ differs from $|\text{MPC}_J\rangle$ by the occupation of two neutron (proton) states:
\begin{equation}
\begin{split}
&\langle\cdots 1_{\zeta,\tau_0}\cdots 1_{\xi,\tau_0}\cdots 0_{\lambda,\tau_0}\cdots 0_{\alpha,\tau_0}\cdots|c_{i,\tau}^{\dagger}c_{j,\tau}^{\dagger}c_{k,\tau}c_{l,\tau}|\cdots 1_{\alpha,\tau_0}\cdots 1_{\lambda,\tau_0} \cdots 0_{\xi,\tau_0}\cdots 0_{\zeta,\tau_0}\rangle\\
=&(-1)^{\mathcal{P}}(-\delta_{i\xi}\delta_{j\zeta}\delta_{l\lambda}\delta_{k\alpha}+\delta_{i\xi}\delta_{j\zeta}\delta_{l\alpha}\delta_{k\lambda}+\delta_{i\zeta}\delta_{j\xi}\delta_{l\lambda}\delta_{k\alpha}-\delta_{i\zeta}\delta_{j\xi}\delta_{l\alpha}\delta_{k\lambda})\delta_{\tau\tau_0},
\end{split}
\end{equation}
where $\mathcal{P}=\sum_{k=1}^{\zeta-1}I_{k,\tau_0}+\sum_{k=1}^{\xi-1}I_{k,\tau_0}+\sum_{k=1}^{\alpha-1}I_{k,\tau_0}^{\prime}+\sum_{k=1}^{\lambda-1}I_{k,\tau_0}^{\prime}$.

\section{Three-body and four-body interaction terms}\label{appendixB}

The three-body and four-body interaction terms in the Hamiltonian \eqref{Hamiltonian} read,
\begin{equation}
\hat{H}_{\text{hot}}=\int d^3r \mathcal{H}^{\text{hot}}=\int d^3r\left\{\frac{1}{3}\beta_S(\bar{\psi}\psi)^3+\frac{1}{4}\gamma_S(\bar{\psi}\psi)^4+\frac{1}{4}\gamma_V\left[(\bar{\psi}\gamma_{\mu}\psi)(\bar{\psi}\gamma^{\mu}\psi)\right]^2\right\}.
\end{equation}

In the second-quantized form, $\hat{H}_{\text{hot}}$ can be written as 
\begin{equation}
\begin{split}
\hat{H}_{\text{hot}}=&\frac{1}{3}\sum_{\tau\tau_1\tau_2}\sum_{\alpha\lambda\alpha_1\lambda_1\alpha_2\lambda_2}D_{\alpha\lambda\alpha_1\lambda_1\alpha_2\lambda_2}^{\tau\tau_1\tau_2}(c_{\alpha,\tau}^{\dagger}c_{\lambda,\tau})(c_{\alpha_1,\tau_1}^{\dagger}c_{\lambda_1,\tau_1})(c_{\alpha_2,\tau_2}^{\dagger}c_{\lambda_2,\tau_2})\\
+&\frac{1}{4}\sum_{\tau\tau_1\tau_2\tau_3}\sum_{\alpha\lambda\alpha_1\lambda_1\alpha_2\lambda_2\alpha_3\lambda_3}K_{\alpha\lambda\alpha_1\lambda_1\alpha_2\lambda_2\alpha_3\lambda_3}^{\tau\tau_1\tau_2\tau_3}(c_{\alpha,\tau}^{\dagger}c_{\lambda,\tau})(c_{\alpha_1,\tau_1}^{\dagger}c_{\lambda_1,\tau_1})(c_{\alpha_2,\tau_2}^{\dagger}c_{\lambda_2,\tau_2})(c_{\alpha_3,\tau_3}^{\dagger}c_{\lambda_3,\tau_3}),
\end{split}
\end{equation}
where
\begin{subequations}
\begin{equation}
D_{\alpha\lambda\alpha_1\lambda_1\alpha_2\lambda_2}^{\tau\tau_1\tau_2}=\int d^3r\left[\beta_S(\bar{\phi}_{\alpha,\tau}\phi_{\lambda,\tau})(\bar{\phi}_{\alpha_1,\tau_1}\phi_{\lambda_1,\tau_1})(\bar{\phi}_{\alpha_2,\tau_2}\phi_{\lambda_2,\tau_2})\right],
\end{equation}
\begin{equation}
\begin{split}
K_{\alpha\lambda\alpha_1\lambda_1\alpha_2\lambda_2\alpha_3\lambda_3}^{\tau\tau_1\tau_2\tau_3}=\int d^3r&\left[\gamma_S(\bar{\phi}_{\alpha,\tau}\phi_{\lambda,\tau})(\bar{\phi}_{\alpha_1,\tau_1}\phi_{\lambda_1,\tau_1})(\bar{\phi}_{\alpha_2,\tau_2}\phi_{\lambda_2,\tau_2})(\bar{\phi}_{\alpha_3,\tau_3}\phi_{\lambda_3,\tau_3})\right.\\
&\left. +\gamma_V(\bar{\phi}_{\alpha,\tau}\gamma_{\mu}\phi_{\lambda,\tau})(\bar{\phi}_{\alpha_1,\tau_1}\gamma^{\mu}\phi_{\lambda_1,\tau_1})(\bar{\phi}_{\alpha_2,\tau_2}\gamma_{\nu}\phi_{\lambda_2,\tau_2})(\bar{\phi}_{\alpha_3,\tau_3}\gamma^{\nu}\phi_{\lambda_3,\tau_3})\right].\\
\end{split}
\end{equation}
\end{subequations}

In the equations of motion for single-particle states, including three-body and four-body interaction terms necessitates adding two additional terms on the right-hand side of Eq. \eqref{EOM-sp-VS}.
\begin{equation}
\sum_{\tau_1\tau_2}\sum_{ji_1j_1i_2j_2}D_{uji_1j_1i_2j_2}^{\tau\tau_1\tau_2}\rho_{iji_1j_1i_2j_2}^{\tau\tau_1\tau_2}+\sum_{\tau_1\tau_2\tau_3}\sum_{ji_1j_1i_2j_2i_3j_3}K_{uji_1j_1i_2j_2i_3j_3}^{\tau\tau_1\tau_2\tau_3}\rho_{iji_1j_1i_2j_2i_3j_3}^{\tau\tau_1\tau_2\tau_3},
\end{equation}
with the three-body and four-body density matrices for the particle-hole channel defined 
\begin{subequations}
\begin{equation}
\label{Appen-high-rho3}
\begin{split}
\rho_{iji_1j_1i_2j_2}^{\tau\tau_1\tau_2}&=\langle\Psi|(c_{i,\tau}^{\dagger}c_{j,\tau})(c_{i_1,\tau_1}^{\dagger}c_{j_1,\tau_1})(c_{i_2,\tau_2}^{\dagger}c_{j_2,\tau_2})|\Psi\rangle\\
&=\sum_{J,I}C_J^*C_I\langle\text{MPC}_J|(c_{i,\tau}^{\dagger}c_{j,\tau})(c_{i_1,\tau_1}^{\dagger}c_{j_1,\tau_1})(c_{i_2,\tau_2}^{\dagger}c_{j_2,\tau_2})|\text{MPC}_I\rangle,
\end{split}
\end{equation}
\begin{equation}
\label{Appen-high-rho4}
\begin{split}
\rho_{iji_1j_1i_2j_2i_3j_3}^{\tau\tau_1\tau_2\tau_3}&=\langle\Psi|(c_{i,\tau}^{\dagger}c_{j,\tau})(c_{i_1,\tau_1}^{\dagger}c_{j_1,\tau_1})(c_{i_2,\tau_2}^{\dagger}c_{j_2,\tau_2})(c_{i_3,\tau_3}^{\dagger}c_{j_3,\tau_3})|\Psi\rangle\\
&=\sum_{J,I}C_J^*C_I\langle\text{MPC}_J|(c_{i,\tau}^{\dagger}c_{j,\tau})(c_{i_1,\tau_1}^{\dagger}c_{j_1,\tau_1})(c_{i_2,\tau_2}^{\dagger}c_{j_2,\tau_2})(c_{i_3,\tau_3}^{\dagger}c_{j_3,\tau_3})|\text{MPC}_I\rangle, \\
\end{split}
\end{equation}
\end{subequations}
respectively. In the equations of motion for the expansion coefficients, the Hamiltonian kernels for the three-body and four-body interaction terms read
\begin{equation}
\label{Appen-high-HJI}
\begin{split}
\mathcal{H}_{JI}^{\text{hot}}=&\langle\text{MPC}_J|\hat{H}_{\text{hot}}|\text{MPC}_I\rangle\\
=&\frac{1}{3}\sum_{\tau\tau_1\tau_2}\sum_{iji_1j_1i_2j_2}D_{iji_1j_1i_2j_2}^{\tau\tau_1\tau_2}\langle\text{MPC}_J|(c_{i,\tau}^{\dagger}c_{j,\tau})(c_{i_1,\tau_1}^{\dagger}c_{j_1,\tau_1})(c_{i_2,\tau_2}^{\dagger}c_{j_2,\tau_2})|\text{MPC}_I\rangle\\
+&\frac{1}{4}\sum_{\tau\tau_1\tau_2\tau_3}\sum_{iji_1j_1i_2j_2i_3j_3}K_{iji_1j_1i_2j_2i_3j_3}^{\tau\tau_1\tau_2\tau_3}\langle\text{MPC}_J|(c_{i,\tau}^{\dagger}c_{j,\tau})(c_{i_1,\tau_1}^{\dagger}c_{j_1,\tau_1})(c_{i_2,\tau_2}^{\dagger}c_{j_2,\tau_2})(c_{i_3,\tau_3}^{\dagger}c_{j_3,\tau_3})|\text{MPC}_I\rangle.\\
\end{split}
\end{equation}

The three- and four-body matrix elements $\langle\text{MPC}_J|(c_{i,\tau}^{\dagger}c_{j,\tau})(c_{i_1,\tau_1}^{\dagger}c_{j_1,\tau_1})(c_{i_2,\tau_2}^{\dagger}c_{j_2,\tau_2})|\text{MPC}_I\rangle$ and $\langle\text{MPC}_J|(c_{i,\tau}^{\dagger}c_{j,\tau})(c_{i_1,\tau_1}^{\dagger}c_{j_1,\tau_1})(c_{i_2,\tau_2}^{\dagger}c_{j_2,\tau_2})(c_{i_3,\tau_3}^{\dagger}c_{j_3,\tau_3})|\text{MPC}_I\rangle$ appearing in Eqs. \eqref{Appen-high-rho3}, \eqref{Appen-high-rho4}, and \eqref{Appen-high-HJI} are computed in a way analogous to the two-body terms in Appendix A, and thus the procedure is not repeated here.

Note that the number of possible Wick contractions for the three-body and four-body terms is much larger than for the two-body terms, which leads to significantly higher computational cost.
Therefore, in the present work, only the cases where $|\text{MPC}_J\rangle$ differs from $|\text{MPC}_I\rangle$ by the occupation of  up to two single-nucleon states are considered for the three‑body and four‑body terms, and the summations in Eqs. \eqref{Appen-high-rho3} and \eqref{Appen-high-rho4} are truncated by retaining only terms with $C_J^*C_I>10^{-6}$.
As shown in Sec. \ref{Numerical-test}, these two approximations have little influence on the conservation of particle number and total energy.

\section{Calculation of $\eta_{ma}^{\tau}$ and $\eta_{am}^{\tau}$}\label{appendixC}

To calculate the anti-Hermitian matrix $\eta_{ma}^{\tau}=\langle\phi_{m,\tau}|\dot{\phi}_{a,\tau}\rangle$, one starts from the original equations of motion for single-particle states [Eq. \eqref{EOM-VS}]  
\begin{equation}
\label{EOM-VS-appendix}
\sum_{\alpha}|\phi_{\alpha,\tau}\rangle\langle\Psi|c_{i,\tau}^{\dagger}c_{\alpha,\tau}\left[-\mathrm{i}\hbar\sum_I\dot{C}_I|\text{MPC}_I\rangle+\left(\hat{H}-\mathrm{i}\hbar\sum_{\tau^{\prime}}\sum_{\xi\zeta}\eta_{\xi\zeta}^{\tau^{\prime}} c_{\xi,\tau^{\prime}}^{\dagger}c_{\zeta,\tau^{\prime}}\right)|\Psi\rangle\right]-\sum_j\mu_{ij}^{\tau}|\phi_{j,\tau}\rangle=0.
\end{equation}
Multiplying Eq. \eqref{EOM-VS-appendix} from the left by a valence state $\langle \phi_{r,\tau}|$, and using the orthogonality of single-particle states, one obtains
\begin{equation}
\langle\Psi|c_{i,\tau}^{\dagger}c_{r,\tau}\left[-\mathrm{i}\hbar\sum_I\dot{C}_I|\text{MPC}_I\rangle+\left(\hat{H}-\mathrm{i}\hbar\sum_{\tau^{\prime}}\sum_{\xi\zeta}\eta_{\xi\zeta}^{\tau^{\prime}} c_{\xi,\tau^{\prime}}^{\dagger}c_{\zeta,\tau^{\prime}}\right)|\Psi\rangle\right]-\mu_{ir}^{\tau}=0.
\end{equation}
Assuming $c_{i,\tau}^{\dagger}$ creates a core state, and the index changes from $i$ to $a$ as defined in Fig. \ref{Division-of-Hilbert-space}, and given that $\langle\Psi|c_{a,\tau}^{\dagger}c_{r,\tau}|\text{MPC}_I\rangle=0$, it follows
\begin{equation}
\label{muam-appendix-1}
\langle\Psi|c_{a,\tau}^{\dagger}c_{r,\tau}\left(\hat{H}-\mathrm{i}\hbar\sum_{\tau^{\prime}}\sum_{\xi\zeta}\eta_{\xi\zeta}^{\tau^{\prime}} c_{\xi,\tau^{\prime}}^{\dagger}c_{\zeta,\tau^{\prime}}\right)|\Psi\rangle-\mu_{ar}^{\tau}=0.
\end{equation}
Taking the Hermitian conjugate of Eq. \eqref{muam-appendix-1}, and then exchanging $a$ and $r$, the following relation is obtained
\begin{equation}
\label{muam-appendix-2}
\langle\Psi|\left(\hat{H}-\mathrm{i}\hbar\sum_{\tau^{\prime}}\sum_{\xi\zeta}\eta_{\xi\zeta}^{\tau^{\prime}} c_{\xi,\tau^{\prime}}^{\dagger}c_{\zeta,\tau^{\prime}}\right)c_{a,\tau}^{\dagger}c_{r,\tau}|\Psi\rangle-(\mu_{ra}^{\tau})^*=0.
\end{equation}
Since $\mu_{ar}^{\tau}=(\mu_{ra}^{\tau})^*$, subtracting Eq. \eqref{muam-appendix-1} from Eq. \eqref{muam-appendix-2} eliminates the Lagrange multipliers,
\begin{equation}
\label{eta-appendix}
\langle\Psi|\left(\hat{H}-\mathrm{i}\hbar\sum_{\tau^{\prime}}\sum_{\xi\zeta}\eta_{\xi\zeta}^{\tau^{\prime}} c_{\xi,\tau^{\prime}}^{\dagger}c_{\zeta,\tau^{\prime}}\right)c_{a,\tau}^{\dagger}c_{r,\tau}|\Psi\rangle-\langle\Psi|c_{a,\tau}^{\dagger}c_{r,\tau}\left(\hat{H}-\mathrm{i}\hbar\sum_{\tau^{\prime}}\sum_{\xi\zeta}\eta_{\xi\zeta}^{\tau^{\prime}} c_{\xi,\tau^{\prime}}^{\dagger}c_{\zeta,\tau^{\prime}}\right)|\Psi\rangle=0.
\end{equation}

As defined in Sec. \ref{Theory}, the nuclear many-body model Hamiltonian in second-quantized form is written as 
\begin{equation}
\label{H-second-quan}
\begin{split}
\hat{H}=&\hat{H}_{\text{DFT}}+\hat{H}_{\text{pair}}\\
=&\sum_{\tau}\sum_{\alpha\lambda}h_{\alpha\lambda}^{\tau}c_{\alpha,\tau}^{\dagger}c_{\lambda,\tau}+\frac{1}{2}\sum_{\tau\tau^{\prime}}\sum_{\alpha\lambda\xi\zeta}W_{\alpha\lambda\xi\zeta}^{\tau\tau^{\prime}}(c_{\alpha,\tau}^{\dagger}c_{\lambda,\tau})(c_{\xi,\tau^{\prime}}^{\dagger}c_{\zeta,\tau^{\prime}})\\
&+\int d^3r\frac{1}{2}A_{\mu}\Delta A^{\mu}+\sum_{\tau}\sum_{\alpha\lambda\xi\zeta}Q_{\alpha\lambda\xi\zeta}^{\tau}c_{\alpha,\tau}^{\dagger}c_{\lambda,\tau}^{\dagger}c_{\xi,\tau}c_{\zeta,\tau}.
\end{split}
\end{equation} 

Substituting the Hamiltonian \eqref{H-second-quan} into Eq. \eqref{eta-appendix} and performing some algebra with the help of Wick’s theorem \cite{Wick1950PR}, we obtain
\begin{equation}
\label{bar-rho-nr}
\sum_{n}i\hbar\bar{\rho}_{nr}^{\tau}\eta_{na}^{\tau}=\sum_n\bar{\rho}_{nr}^{\tau}h_{na}^{\tau}+\sum_{jkl}\tilde{Q}_{rjkl}^{\tau}\kappa_{ajkl}^{\tau}-\sum_{ijk}\bar{Q}_{ijka}^{\tau}\kappa_{ijkr}^{\tau}+\sum_{\tau^{\prime}}\sum_{jkl}W_{rjkl}^{\tau\tau^{\prime}}\rho_{ajkl}^{\tau\tau^{\prime}}-\sum_{\tau^{\prime}}\sum_{ikl}W_{iakl}^{\tau\tau^{\prime}}\rho_{irkl}^{\tau\tau^{\prime}},
\end{equation}
with $\bar{\rho}_{nr}^{\tau}=\delta_{nr}-\rho_{nr}^{\tau}$, and $\bar{Q}_{ijka}^{\tau}=Q_{ijka}^{\tau}-Q_{ijak}^{\tau}$. 

Multiplying both sides of Eq. \eqref{bar-rho-nr} by $\sum_{r}\bar{\rho}_{rm}^{-1}$ yields
\begin{equation}
i\hbar\eta_{ma}^{\tau}=h_{ma}^{\tau}+\sum_r\left(\sum_{jkl}\tilde{Q}_{rjkl}^{\tau}\kappa_{ajkl}^{\tau}-\sum_{ijk}\bar{Q}_{ijka}^{\tau}\kappa_{ijkr}^{\tau}+\sum_{\tau^{\prime}}\sum_{jkl}W_{rjkl}^{\tau\tau^{\prime}}\rho_{ajkl}^{\tau\tau^{\prime}}-\sum_{\tau^{\prime}}\sum_{ikl}W_{iakl}^{\tau\tau^{\prime}}\rho_{irkl}^{\tau\tau^{\prime}}\right)\bar{\rho}_{rm}^{-1}.
\end{equation}
This is the formula for calculating $\eta_{ma}^{\tau}$. In an analogous way, 
$\eta_{am}^{\tau}$ is calculated.

\section{RDFT-SLAP framework}\label{appendixD}

In principle, the initial wave function of CI-TDDFT can be chosen as the ground state of the many-body Hamiltonian \eqref{Hamiltonian}. The ground state can be obtained by an imaginary-time evolution of the equations of motion for the single-particle states \eqref{phi-a}, \eqref{phi-r} and expansion coefficients \eqref{EOM-C}.
However, because of the problem of variational collapse in the case of relativistic density functionals \cite{Zhang2010IJMPE_TsunamiDiracSea,Tanimura2015PTEP_3DMeshCDFT,RenZX2017PRC}, we cannot directly employ the imaginary-time evolution to obtain the ground state, and thus have to use an alternative method to specify the initial wave function.

The RDFT-SLAP method takes into account pairing correlations and blocking effects by diagonalizing the pairing Hamiltonian in a truncated MPC space with good particle number \cite{MengJ2006FPC}.
The form of the RDFT-SLAP ground-state wave function is the same as in Eq. \eqref{wave-function} and, 
therefore, it can be used as the initial state for CI-TDDFT.
The detailed formalism of RDFT-SLAP can be found in Refs. \cite{MengJ2006FPC,ShiZ2018PRC,WangYP2023PLB}. Here we only include a brief outline.

In RDFT-SLAP, the model Hamiltonian reads
\begin{equation}
\hat{H}_{\text{RDFT-SLAP}}=\hat{H}_{\text{pair}}+\hat{H}_0,
\end{equation}
where $\hat{H}_{\text{pair}}$ is the pairing term, and $\hat{H}_0$ is the one-body Hamiltonian.
For the pairing interaction, a monopole pairing Hamiltonian as defined in Eq. \eqref{pairing-Hamiltonian} is adopted. 
The one-body term is given by
\begin{equation}
\hat{H}_0=\sum_{\tau}\sum_{i}\varepsilon_{i,\tau}c_{i,\tau}^{\dagger}c_{i,\tau},
\end{equation}
where $c_{i,\tau}^{\dagger}$ is the creation operator of the single-particle state $|\phi_{i,\tau}\rangle$, and $\varepsilon_{i,\tau}$ is the corresponding single-particle energy obtained from the Dirac equation,
\begin{equation}
\label{Dirac}
\left[\boldsymbol{\alpha}(-\text{i}\boldsymbol{\nabla}-\boldsymbol{V})+\beta(m+S)+V^0\right]\phi_{i,\tau}=\varepsilon_{i,\tau}\phi_{i,\tau}.
\end{equation}
Here, the scalar field $S$ and vector field $V^{\mu}$ are respectively given by
\begin{subequations}
\begin{equation}
\label{S-field}
S=\alpha_S\rho_S+\beta_S\rho_S^2+\gamma_S\rho_S^3+\delta_S\Delta\rho_S,
\end{equation}
\begin{equation}
\label{V-field}
V^{\mu}=\alpha_Vj^{\mu}+\gamma_V(j_{\nu}j^{\nu})j^{\mu}+\delta_V\Delta j^{\mu}+\tau_3\alpha_{TV}j_{TV}^{\mu}+\tau_3\delta_{TV}\Delta j_{TV}^{\mu}+e\frac{1-\tau_3}{2}A^{\mu},
\end{equation}
\end{subequations}
where $\rho_S$, $j^{\mu}$, and $j_{TV}^{\mu}$ are local densities and currents. 
In our case, specifically, to start monopole density oscillations of $^{58}$Ni and $^{60}$Ni, an additional constraining potential is included in the Dirac equation \eqref{Dirac} during the RDFT-SLAP calculation to obtain a ground state with a chosen radius.
The standard augmented Lagrangian method \cite{Staszczak2010EPJA} is used.
Note that this constraining potential does not appear in the subsequent CI‑TDDFT time evolution.

The RDFT-SLAP wave function has the same form as in Eq. \eqref{wave-function},
\begin{equation}
|\Psi(t_0)\rangle = \sum_I C_I(t_0) |\text{MPC}_I(t_0)\rangle.
\end{equation}
The expansion coefficients $C_I(t_0)$ are obtained by diagonalizing the Hamiltonian $\hat{H}_{\text{RDFT-SLAP}}$ in the MPC space. The resulting $C_I(t_0)$ determine the occupation probabilities $n_{i,\tau}$ of single-particle states [Eq. \eqref{n-i-tau}], and the latter are then used to update the local densities and currents, 
\begin{subequations}
\begin{equation}
\rho_S=\sum_{\tau}\sum_{i} n_{i,\tau}\bar{\phi}_{i,\tau}\phi_{i,\tau},
\end{equation}
\begin{equation}
j^{\mu}=\sum_{\tau}\sum_{i} n_{i,\tau}\bar{\phi}_{i,\tau}\gamma^{\mu}\phi_{i,\tau},
\end{equation}
\begin{equation}
j_{TV}^{\mu}=\sum_{\tau}\sum_{i} n_{i,\tau}\bar{\phi}_{i,\tau}\gamma^{\mu}\tau_3\phi_{i,\tau},
\end{equation}
\end{subequations}
which, in turn, define the scalar field $S$ \eqref{S-field} and the vector field $V^{\mu}$ \eqref{V-field}.
Therefore, the RDFT-SLAP equations are solved iteratively untill self-consistency is achieved.

The final self-consistent RDFT-SLAP wave function is used as the initial state for CI-TDDFT. 
Although this wave function is not an exact ground state of the many-body Hamiltonian \eqref{Hamiltonian}, it captures the essential short-range pairing correlations and long-range particle-hole correlations, and has  proven effective in describing nuclear ground-state properties \cite{MengJ2006FPC,WangYP2024PRL} as well as rotational properties \cite{ShiZ2018PRC,WangYP2023PLB,XuFF2024PRL}.

\section{Strength function}\label{appendixD}

The strength function is obtained from the time-dependent expectation value of the monopole operator following standard linear-response theory \cite{Ring2004The,Nakatsukasa2016RMP}.
In this appendix, units with $\hbar=1$ are used.
To model small-amplitude monopole oscillation, we start from a state with a radius $r$ different from that of the ground state. 
The constraining potential reads
\begin{equation}
V_{\text{constr}} = \lambda F,\quad F=r^2,
\end{equation}
where $\lambda$ is the constraint parameter.

Let us assume that a nucleus is in its ground state $|\Phi_0\rangle$, with energy $E_0=0$ at $t=-\infty$, and the external constraining field is adiabatically switched on from $t=-\infty$ to $t=0$,
\begin{equation}
V_{\text{ext}}^{\prime}(t)=\text{lim}_{\epsilon\rightarrow 0}V_{\text{ext}}(t)e^{\epsilon t},\qquad V_{\text{ext}}(t)\equiv V_{\text{constr}}(t) = \lambda F \theta(-t),
\end{equation}
where $\theta(t)$ denotes the Heaviside step function.

The field $V_{\text{ext}}(t)$ can be expressed in terms of a Fourier transform:
\begin{equation}
V_{\text{ext}}(t)=\frac{1}{2\pi}\int_0^{\infty}[V_{\text{ext}}(\omega)Fe^{-i\omega t}+V_{\text{ext}}^*(\omega)F^{\dagger}e^{i\omega t}]d\omega.
\end{equation}
Accordingly, $V_{\text{ext}}^{\prime}(t)$ reads
\begin{equation}
V_{\text{ext}}^{\prime}(t)=\text{lim}_{\epsilon\rightarrow 0}\frac{1}{2\pi}\int_0^{\infty}[V_{\text{ext}}(\omega)Fe^{-i(\omega+i\epsilon) t}+V_{\text{ext}}^*(\omega)F^{\dagger}e^{i(\omega-i\epsilon) t}]d\omega.
\end{equation}
Using time-dependent perturbation theory, the wave function influenced by the external field $V^{\prime}_{\text{ext}}(t)$ at time $t$ can be expressed in the first-order approximation as 
\begin{equation}
|\Psi(t)\rangle=|\Phi_0\rangle-i\sum_ne^{-iE_nt}\int_{-\infty}^tdt^{\prime}e^{iE_nt^{\prime}}|\Phi_n\rangle\langle\Phi_n|V_{\text{ext}}^{\prime}(t^{\prime})|\Phi_0\rangle,
\end{equation}
that is,
\begin{equation}
|\Psi(t)\rangle=|\Phi_0\rangle+\sum_n|\Phi_n\rangle\times \text{lim}_{\epsilon\rightarrow 0}\frac{1}{2\pi}\int_0^{\infty}\left[\frac{V_{\text{ext}}(\omega)\langle\Phi_n|F|\Phi_0\rangle}{\omega-E_n+i\epsilon}\times e^{-i(\omega+i\epsilon)t}-\frac{V_{\text{ext}}^*(\omega)\langle\Phi_n|F^{\dagger}|\Phi_0\rangle}{\omega+E_n-i\epsilon}e^{i(\omega-i\epsilon)t}\right]d\omega.
\end{equation}
The time-dependent expectation value of an operator $F$ is defined with the following expression 
\begin{equation}
\label{Ft1}
\begin{split}
F(t)=&\langle\Psi(t)|F^{\dagger}|\Psi(t)\rangle-\langle\Phi_0|F^{\dagger}|\Phi_0\rangle\\
=&\sum_n\langle\Phi_0|F^{\dagger}|\Phi_n\rangle\times \text{lim}_{\epsilon\rightarrow 0}\frac{1}{2\pi}\int_{0}^{\infty}\left[\frac{V_{\text{ext}}(\omega)\langle\Phi_n|F|\Phi_0\rangle}{\omega-E_n+i\epsilon}\times e^{-i(\omega+i\epsilon)t}-\frac{V_{\text{ext}}^*(\omega)\langle\Phi_n|F^{\dagger}|\Phi_0\rangle}{\omega+E_n-i\epsilon}e^{i(\omega-i\epsilon)t}\right]d\omega\\
&+\sum_n\langle\Phi_n|F^{\dagger}|\Phi_0\rangle\times \text{lim}_{\epsilon\rightarrow 0}\frac{1}{2\pi}\int_{0}^{\infty}\left[\frac{V_{\text{ext}}^*(\omega)\langle\Phi_0|F^{\dagger}|\Phi_n\rangle}{\omega-E_n-i\epsilon}\times e^{i(\omega-i\epsilon)t}-\frac{V_{\text{ext}}(\omega)\langle\Phi_0|F|\Phi_n\rangle}{\omega+E_n+i\epsilon}e^{-i(\omega+i\epsilon)t}\right]d\omega\\
=&\text{lim}_{\epsilon\rightarrow 0}\frac{1}{2\pi}\int_0^{\infty}V_{\text{ext}}(\omega)R_F(\omega)e^{-i(\omega+i\epsilon)t}d\omega+c.c.,
\end{split}
\end{equation}
where $R_F(\omega)$ reads:
\begin{equation}
R_F(\omega)=\sum_n\left(\frac{|\langle\Phi_n|F|\Phi_0\rangle|^2}{\omega-E_n+i\epsilon}-\frac{|\langle\Phi_n|F^{\dagger}|\Phi_0\rangle|^2}{\omega+E_n+i\epsilon}\right).
\end{equation}
The time evolution of $F(t)$ can also be expressed in terms of a Fourier transformation:
\begin{equation}
\label{Ft2}
F(t)=\frac{1}{2\pi}\int_0^{\infty}[F(\omega)e^{-i\omega t}+F^*(\omega)e^{i\omega t}]d\omega.
\end{equation}
Comparing the two expressions \eqref{Ft1} and \eqref{Ft2} in the limit $\epsilon\rightarrow 0$, one obtains for $R_F(\omega)$:
\begin{equation}
R_F(\omega)=\frac{F(\omega)}{V_{\text{ext}}(\omega)}.
\end{equation}
The Fourier transform of $V_{\text{ext}}(t)$ yields $V_{\text{ext}}(\omega)$
\begin{equation}
V_{\text{ext}}(\omega)=\text{lim}_{\epsilon\rightarrow 0}\int_{-\infty}^{+\infty}\left[\lambda \theta(-t)e^{i(\omega-i\epsilon)t}\right]dt=\frac{\lambda}{i\omega}, \qquad \omega>0,
\end{equation} 
and, therefore,
\begin{equation}
R_F(\omega)=\frac{i\omega F(\omega)}{\lambda}, \qquad \omega>0.
\end{equation}
Finally, the strength function is given by,
\begin{equation}
\label{Somega}
S(\omega)=-\frac{1}{\pi}\text{Im}R_F(\omega)=-\frac{1}{\pi}\text{Im}\frac{i\omega F(\omega)}{\lambda},\qquad \omega>0.
\end{equation}

\date{today}

\newpage

\begin{acknowledgments}

This work was partly supported by the National Natural Science Foundation of China (12505135, 12435006, 12475117, 12141501), the High-End Foreign Experts Plan of China, the National Key Laboratory of Neutron Science and Technology (Grant No. NST202401016), the National Key R\&D Program of China 2024YFE0109803, the National Key Research and Development Program of China 2024YFA1612600, the High-performance Computing Platform of Peking University, and the China Postdoctoral Science Foundation (2025M773401), the project “Implementation of cutting-edge research and its application as part of the Scientific Center of Excellence for Quantum and Complex Systems, and Representations of Lie Algebras", Grant No. PK.1.1.10.0004,
co-financed by the European Union through the European Regional Development Fund -
Competitiveness and Cohesion Programme 2021-2027; and the Croatian Science Foundation under the project Dynamics of complex femtosystems (IP-2025-02-4214).

\end{acknowledgments}

\end{document}